\documentclass[iop,revtex4-1]{emulateapj}

\usepackage{natbib}
\usepackage{hyperref}
\usepackage{amsmath}
\usepackage{xspace}
\usepackage{lineno}

\usepackage{multirow}
\usepackage{threeparttable}

\newcommand{\degree}{\ensuremath{{}^{\circ}}\xspace}

\mathchardef\mhyphen="2D
\graphicspath{{./fig/}}

\shorttitle{Chromatic Errors that Impact Sub--1\% Photometry}
\shortauthors{Li et al.}
\begin{document}

\title{Assessment of Systematic Chromatic Errors that Impact Sub--1\% Photometric Precision in Large-Area Sky Surveys}

\author{
T.~S.~Li\altaffilmark{1},
D.~L.~DePoy\altaffilmark{1},
J.~L.~Marshall\altaffilmark{1},
D.~Tucker\altaffilmark{2},
R.~Kessler\altaffilmark{3},
J.~Annis\altaffilmark{2},
G.~M.~Bernstein\altaffilmark{4},
S.~Boada\altaffilmark{1},
D.~L.~Burke\altaffilmark{5,6},
D.~A.~Finley\altaffilmark{2},
D.~J.~James\altaffilmark{7},
S.~Kent\altaffilmark{2},
H.~Lin\altaffilmark{2},
J.~Marriner\altaffilmark{2},
N.~Mondrik\altaffilmark{1},
D.~Nagasawa\altaffilmark{1},
E.~S.~Rykoff\altaffilmark{6,5},
D.~Scolnic\altaffilmark{3},
A.~R.~Walker\altaffilmark{7},
W.~Wester\altaffilmark{2},
T. M. C.~Abbott\altaffilmark{7},
S.~Allam\altaffilmark{2},
A.~Benoit-L{\'e}vy\altaffilmark{8,9,10},
E.~Bertin\altaffilmark{8,10},
D.~Brooks\altaffilmark{9},
D.~Capozzi\altaffilmark{11},
A.~Carnero~Rosell\altaffilmark{12,13},
M.~Carrasco~Kind\altaffilmark{14,15},
J.~Carretero\altaffilmark{16,17},
M.~Crocce\altaffilmark{16},
C.~E.~Cunha\altaffilmark{6},
C.~B.~D'Andrea\altaffilmark{11,18},
L.~N.~da Costa\altaffilmark{12,13},
S.~Desai\altaffilmark{19,20},
H.~T.~Diehl\altaffilmark{2},
P.~Doel\altaffilmark{9},
B.~Flaugher\altaffilmark{2},
P.~Fosalba\altaffilmark{16},
J.~Frieman\altaffilmark{2,3},
E.~Gaztanaga\altaffilmark{16},
D.~A.~Goldstein\altaffilmark{21,22},
D.~Gruen\altaffilmark{6,23,5,24},
R.~A.~Gruendl\altaffilmark{14,15},
G.~Gutierrez\altaffilmark{2},
K.~Honscheid\altaffilmark{25,26},
K.~Kuehn\altaffilmark{27},
N.~Kuropatkin\altaffilmark{2},
M.~A.~G.~Maia\altaffilmark{12,13},
P.~Melchior\altaffilmark{28},
C.~J.~Miller\altaffilmark{29,30},
R.~Miquel\altaffilmark{31,17},
J.~J.~Mohr\altaffilmark{19,20,23},
E.~Neilsen\altaffilmark{2},
R.~C.~Nichol\altaffilmark{11},
B.~Nord\altaffilmark{2},
R.~Ogando\altaffilmark{12,13},
A.~A.~Plazas\altaffilmark{32},
A.~K.~Romer\altaffilmark{33},
A.~Roodman\altaffilmark{6,5},
M.~Sako\altaffilmark{4},
E.~Sanchez\altaffilmark{34},
V.~Scarpine\altaffilmark{2},
M.~Schubnell\altaffilmark{30},
I.~Sevilla-Noarbe\altaffilmark{34,14},
R.~C.~Smith\altaffilmark{7},
M.~Soares-Santos\altaffilmark{2},
F.~Sobreira\altaffilmark{2,12},
E.~Suchyta\altaffilmark{4},
G.~Tarle\altaffilmark{30},
D.~Thomas\altaffilmark{11},
V.~Vikram\altaffilmark{35}
\\ \vspace{0.2cm} (The DES Collaboration) \\
}
 
\altaffiltext{1}{George P. and Cynthia Woods Mitchell Institute for Fundamental Physics and Astronomy, and Department of Physics and Astronomy, Texas A\&M University, College Station, TX 77843,  USA}
\altaffiltext{2}{Fermi National Accelerator Laboratory, P. O. Box 500, Batavia, IL 60510, USA}
\altaffiltext{3}{Kavli Institute for Cosmological Physics, University of Chicago, Chicago, IL 60637, USA}
\altaffiltext{4}{Department of Physics and Astronomy, University of Pennsylvania, Philadelphia, PA 19104, USA}
\altaffiltext{5}{SLAC National Accelerator Laboratory, Menlo Park, CA 94025, USA}
\altaffiltext{6}{Kavli Institute for Particle Astrophysics \& Cosmology, P. O. Box 2450, Stanford University, Stanford, CA 94305, USA}
\altaffiltext{7}{Cerro Tololo Inter-American Observatory, National Optical Astronomy Observatory, Casilla 603, La Serena, Chile}
\altaffiltext{8}{CNRS, UMR 7095, Institut d'Astrophysique de Paris, F-75014, Paris, France}
\altaffiltext{9}{Department of Physics \& Astronomy, University College London, Gower Street, London, WC1E 6BT, UK}
\altaffiltext{10}{Sorbonne Universit\'es, UPMC Univ Paris 06, UMR 7095, Institut d'Astrophysique de Paris, F-75014, Paris, France}
\altaffiltext{11}{Institute of Cosmology \& Gravitation, University of Portsmouth, Portsmouth, PO1 3FX, UK}
\altaffiltext{12}{Laborat\'orio Interinstitucional de e-Astronomia - LIneA, Rua Gal. Jos\'e Cristino 77, Rio de Janeiro, RJ - 20921-400, Brazil}
\altaffiltext{13}{Observat\'orio Nacional, Rua Gal. Jos\'e Cristino 77, Rio de Janeiro, RJ - 20921-400, Brazil}
\altaffiltext{14}{Department of Astronomy, University of Illinois, 1002 W. Green Street, Urbana, IL 61801, USA}
\altaffiltext{15}{National Center for Supercomputing Applications, 1205 West Clark St., Urbana, IL 61801, USA}
\altaffiltext{16}{Institut de Ci\`encies de l'Espai, IEEC-CSIC, Campus UAB, Carrer de Can Magrans, s/n,  08193 Bellaterra, Barcelona, Spain}
\altaffiltext{17}{Institut de F\'{\i}sica d'Altes Energies (IFAE), The Barcelona Institute of Science and Technology, Campus UAB, 08193 Bellaterra (Barcelona) Spain}
\altaffiltext{18}{School of Physics and Astronomy, University of Southampton,  Southampton, SO17 1BJ, UK}
\altaffiltext{19}{Excellence Cluster Universe, Boltzmannstr.\ 2, 85748 Garching, Germany}
\altaffiltext{20}{Faculty of Physics, Ludwig-Maximilians University, Scheinerstr. 1, 81679 Munich, Germany}
\altaffiltext{21}{Department of Astronomy, University of California, Berkeley,  501 Campbell Hall, Berkeley, CA 94720, USA}
\altaffiltext{22}{Lawrence Berkeley National Laboratory, 1 Cyclotron Road, Berkeley, CA 94720, USA}
\altaffiltext{23}{Max Planck Institute for Extraterrestrial Physics, Giessenbachstrasse, 85748 Garching, Germany}
\altaffiltext{24}{Universit\"ats-Sternwarte, Fakult\"at f\"ur Physik, Ludwig-Maximilians Universit\"at M\"unchen, Scheinerstr. 1, 81679 M\"unchen, Germany}
\altaffiltext{25}{Center for Cosmology and Astro-Particle Physics, The Ohio State University, Columbus, OH 43210, USA}
\altaffiltext{26}{Department of Physics, The Ohio State University, Columbus, OH 43210, USA}
\altaffiltext{27}{Australian Astronomical Observatory, North Ryde, NSW 2113, Australia}
\altaffiltext{28}{Department of Astrophysical Sciences, Princeton University, Peyton Hall, Princeton, NJ 08544, USA}
\altaffiltext{29}{Department of Astronomy, University of Michigan, Ann Arbor, MI 48109, USA}
\altaffiltext{30}{Department of Physics, University of Michigan, Ann Arbor, MI 48109, USA}
\altaffiltext{31}{Instituci\'o Catalana de Recerca i Estudis Avan\c{c}ats, E-08010 Barcelona, Spain}
\altaffiltext{32}{Jet Propulsion Laboratory, California Institute of Technology, 4800 Oak Grove Dr., Pasadena, CA 91109, USA}
\altaffiltext{33}{Department of Physics and Astronomy, Pevensey Building, University of Sussex, Brighton, BN1 9QH, UK}
\altaffiltext{34}{Centro de Investigaciones Energ\'eticas, Medioambientales y Tecnol\'ogicas (CIEMAT), Madrid, Spain}
\altaffiltext{35}{Argonne National Laboratory, 9700 South Cass Avenue, Lemont, IL 60439, USA}

\begin{abstract}
Meeting the science goals for many current and future ground-based optical large-area sky surveys requires that the calibrated broadband photometry is both stable in time and uniform over the sky to 1\% precision or better. Past and current surveys have achieved photometric precision of 1--2\% by calibrating the survey's stellar photometry with repeated measurements of a large number of stars observed in multiple epochs.  The calibration techniques employed by these surveys only consider the relative frame-by-frame photometric zeropoint offset and the focal plane position-dependent illumination corrections, which are independent of the source color. However, variations in the wavelength dependence of the atmospheric transmission and the instrumental throughput induce source color-dependent systematic errors. These systematic errors must also be considered to achieve the most precise photometric measurements. In this paper, we examine such systematic chromatic errors using photometry from the Dark Energy Survey (DES) as an example. We first define a natural magnitude system for DES and calculate the systematic errors on stellar magnitudes, when the atmospheric transmission and instrumental throughput deviate from the natural system. 
We conclude that the systematic chromatic errors caused by the change of airmass in each exposure, the change of the precipitable water vapor and aerosol in the atmosphere over time, and the non-uniformity of instrumental throughput over the focal plane, can be up to 2\% in some bandpasses. We then compare the calculated systematic chromatic errors with the observed DES  data. For the test sample data, we correct these errors using measurements of the atmospheric transmission and instrumental throughput from auxiliary calibration systems. The residual after correction is less than 0.3\%. Moreover, we calculate such systematic chromatic errors for Type Ia supernovae and elliptical galaxies and find that the chromatic errors for non-stellar objects are redshift-dependent and can be larger than those for stars at certain redshifts.
\end{abstract}

\section{INTRODUCTION}\label{sec:intro}

In traditional astronomical photometry, a set of standard stars, such as those from Landolt's \citep{Landolt1992} or Stetson's \citep{Stetson2000, Stetson2005} catalogs, is observed over a wide range of airmasses during the course of a night to calibrate all sources observed on the same night. The observed instrumental magnitude $m_b$ and the standard magnitude $m_b^0$ for a given bandpass $b$ have the following relation,
\begin{equation}\label{eq:photocal}
m_b - m_b^0 = a_b + k_b \cdot X +  c_b \cdot color
\end{equation} 
where $a_b$ is the photometric zeropoint, $k_b$ is the first order atmospheric extinction coefficient, $X$ is the airmass for each exposure, $c_b$ is the color term coefficient, and $color$ is the color of the stars (e.g. $g-r$ or $V-R$, depending on the photometric systems and the filter bandpasses). On a photometric night -- i.e., a night in which the atmospheric extinction coefficient is stable over time and uniform over the sky -- the nightly $a_b$ and $k_b$ are derived. The color term $c_b$ is a first-order correction to compensate for the difference in the $shape$ of the filter bandpass of the standard system and that of the filter bandpass actually used in the night's observations. This term corrects for the full system response for that filter bandpass, including both the instrument throughput and the atmospheric transmission.  Fortunately, for most optical passbands, the color term coefficients are reasonably constant over the course of a typical observing run ($\lesssim$1 week), and by fitting the above relation to observations of standard stars and applying the results to the science exposures, all the program target objects can be calibrated to a standard photometric system with reasonable precision.

We note that tying data to a standard system serves two aspects of photometric calibration:  {\em relative} calibration and {\em absolute} calibration.  Relative calibration refers to creating a data set whose photometry is internally consistent:  e.g., the measured brightness and color of a (non-variable) star is, all else being equal, independent of its time of observation or its location on the sky.  Absolute calibration refers to creating a data set whose photometry can be tied to physical units of specific flux \citep[see~e.g.][for a review]{Scolnic2015}.  By tying data to a standard system, one ensures that the data are consistent with the standard photometric system and can connect the apparent brightnesses and colors of stars in one's own data to those of stars that have calibrated magnitudes that are convertible to units of specific flux in ergs~s$^{-1}$~cm$^{-2}$~Hz$^{-1}$~\citep{Holberg2006}.  



In recent years, large imaging surveys like the Sloan Digital Sky Survey (SDSS) have opted to create their own standard photometric systems -- ones based on the ``natural'' photometric system of their instruments\footnote{A natural system is one in which the color term coefficients $c_b$ are all identically zero.  Since system responses can and do change with time or even spatially across a survey instrument's field-of-view, surveys tend to {\em define} their natural systems by their instrument's mean system response, thus ensuring any color terms are very small and average to zero.} -- rather than try to transform their immense quantities of data to a previously defined standard system, like the Johnson-Cousins $UBVR_cIc$ system \citep{Bessell1990,Bessell2012}.  A practical advantage of this is that it effectively decouples the photometric calibration of data taken in one filter with data taken in another filter:  in other words,  one need not match data from one filter to data in another filter in order to apply a color term, and this works sufficiently well for calibrating large optical imaging surveys at the $\sim$2\% (0.02~mag) level.



With the success of these earlier surveys, photometric calibration has become an important factor in the systematic error budgets in the era of precision cosmology. Therefore, many current and future ground-based wide-field imaging surveys have the ambitious calibration goal of ``breaking the 1\% barrier", which requires that calibrated broadband photometry is both stable in time and uniform over the sky to $<1$\% (0.01~mag or 10~millimag) rms precision. These sub--1\% precision requirements are driven by the specific science needs of photometric redshift accuracy, the separation of stellar populations, detection of low-amplitude variable objects, and the search for systematic effects in Type Ia supernova light
curves~\citep[see more details at e.g., the Large Synoptic Survey Telescope Science Book,][]{LSST2009}. 

Traditionally, photometry with $\sim$1\% level precision is reachable when careful analysis is taken on the flat fielding, such as star flats~\citep{Manfroid1995}, across a small field-of-view (FOV), and when observations are done under photometric conditions, i.e. atmospheric conditions are stable and free of clouds. Thanks to the continuous and rapid observing cadence of these dedicated surveys, overlapping areas with  multi-epoch observations can be used to calibrate the illumination pattern of the imaging system with a large FOV. Indeed, Padmanabhan et al. (2008) applied the ``Ubercal" procedure to the SDSS data taken in good photometric conditions and reached rms of 1-2\% relative photometry. Many other sky surveys, such as Pan-STARRS \citep{Schlafly2012} and Deep Lens Survey \citep{Wittman2012} have also adopted this calibration procedure for their photometric calibrations. 

In order to maximize survey efficiency, imaging surveys might also be conducted in less than ideal conditions, i.e. cloudy or partly cloudy nights. Again, owing to the overlapping area in multi-epoch observations, the repeated measurements of a large number of stars allow an internal global calibration of the frame-to-frame zeropoint offset \citep{Glazebrook1994}, which links the instrumental magnitude and natural magnitude of the survey. This zeropoint offset can be a combination of the instrumental zeropoint change, the atmospheric extinction at a given airmass, and cloud extinction. One zeropoint offset is computed and applied to each exposure or each CCD detector, depending on the airmass of the exposure as well as photometric condition. For example, when the airmass is small and the night is photometric, the zeropoint offset could be computed on an exposure-by-exposure level; when the airmass is large or the night is cloudy, then the zeropoint offset could be computed on a CCD-by-CCD level. \cite{MacDonald2004} used this technique on the global calibration of Oxford-Dartmouth Thirty Degree Survey. 

The Dark Energy Survey (DES) is a wide-area photometric survey of 5,000 square degrees using the Dark Energy Camera (DECam) at the Cerro Tololo Inter-American Observatory (CTIO) 4m Blanco telescope. DECam is composed of 74 $250\mu m$ thick fully-depleted CCDs -- 62 for science imaging, plus 12 CCDs for real-time guiding and focus -- with a FOV of 2.2$\degree$ (3.1 deg$^2$ in area) and a pixel scale of 0.26~arcsec/pixel.  
The filters are 620~mm in diameter and fully cover the 62 science CCDs.  For further details on DECam itself, see~\citet{Flaugher2015}. DES has a requirement for relative photometric calibrations: the survey calibrations must be internally consistent both spatially over the survey footprint and temporally over the 5 years of the survey to at least 2\% with a goal of 1\% or better. However, one of its four main probes of cosmological parameters -- the Hubble Diagram of Type Ia supernovae -- requires photometric precision better than 1\% for the 10 supernova fields.   To achieve these relative calibration requirements, DES uses a combination of calibration methods mentioned above. First, star flats are obtained at the beginning of each DES season and during engineering nights in order to obtain robust pupil ghost and illumination corrections for the flat-fielding exposures (G.\ Bernstein et al., in prep.).  Second, over its 5-year run, DES will cover its full footprint 10 times (in 10 ``tilings'') in each of its 5 filters (DES-$grizY$\footnote{DECam has seven filters. They are DECam-$u$, DES-$g$, DES-$r$, DES-$i$, DES-$z$, DES-$Y$, and DECam-$VR$. DES has no $u$-band component in its primary survey. However, we include discussion of DECam-$u$ in this paper since it is available for all DECam community users. For simplicity, we will refer to the six bands as $ugrizY$ in the paper.}), and it uses the large overlaps between exposures in different tilings to tie together the relative calibrations globally across the full survey footprint.
Meanwhile, a sparse grid-work of stars extracted from the multiple DES tilings and calibrated via nightly DES standard star solutions serves both to ``anchor" the relative calibrations against large-scale (but low-amplitude) systematic gradients that are often inherent to Ubercal techniques and to tie the relative calibrations to an absolute flux calibration (\citealt{Tucker2007}; Tucker et al., in prep.).

However, most of the calibration techniques discussed above  consider only the relative frame-by-frame zeropoint offset and position dependent illumination corrections, which are independent of the source color, i.e. grey-scale zeropoint corrections, or grey-term. In reality, variations in the wavelength dependence of the system response (i.e. atmospheric transmission + instrumental throughput) can also induce changes in measuring the brightness of an object that depend on the spectral energy distributions (SEDs) of the object. We refer to such changes as systematic chromatic errors, or SCE, throughout this paper. We use the word ``chromatic", since this effect could be considered as approximately linear to stellar colors, which is similar to a linear color-term correction (e.g. $c_b$ in Equation \ref{eq:photocal}) used to transform from one photometric system to another. It is essentially the change of the $shape$ of the system response. At 1\% level photometric precision, SCE are significant components of the total photometric error budget when calibration techniques only include grey-scale zeropoint corrections. In a few previous imaging surveys, SCE have been partially considered. For example, \cite{Ivezic2007} applied color-term corrections for different transmission curves from six camera columns when making the SDSS standard star catalog for Stripe 82. The Supernova Legacy Survey (SNLS) built a photometric response map to correct  the non-negligible color-term variations between photometric measurements obtained at different focal plane positions of the wide field imager MegaCam~\citep{Regnault2009}. \cite{Betoule2013} also considered such color-term variations in a combined photometric calibration of the SNLS and the SDSS supernova survey. Most past surveys, however, did not include corrections for SCE in their photometric calibration, especially the SCE from the atmospheric variation, as these corrections are small and hard to determine using the data alone. In this paper, we calculate the SCE and show that these errors are caused by not only the non-uniformity of system response function over the focal plane, but also the change of airmass in each exposure and the change of the precipitable water vapor and aerosol in the atmosphere over time. We also demonstrate that our calculations match what we observe in the DES data.

This paper will only discuss the photometric calibrations from the detectors to the top of Earth's atmosphere. It is worth noting that Galactic interstellar extinction is also a very important aspect in order to achieve sub-1\% photometric precision, as the reddening will affect the color of objects measured at the top of the Earth atmosphere. Photometric calibration performed
using the stellar locus regression technique~\citep{Ivezic2004, MacDonald2004, High2009} corrects the zeropoint variation caused by Galactic extinction. Recently, \cite{Yuan2015} used a spectroscopy-based stellar color regression method to reanalyze the SDSS data with spectra obtained from the LAMOST survey~\citep{Deng2012} and delivered an accuracy of a few millimag for color calibration. It is true that interstellar extinction will complicate the uniformity of the zeropoint calibration across the sky. However, the effect of interstellar extinction is somewhat different from the SCE discussed in this paper, since at any given line-of-sight the reddening is constant and should not change the color of objects in repeated observations.

We structure the paper as follows: In Section 2, we discuss possible variations in the system response and define a fiducial system response for the DES natural system. In Section 3, we calculate the synthetic SCE for stellar objects when the system response deviates from a fiducial system response. In Section 4, we compare these synthetic SCE with actual DES data and show the SCE could be corrected using the synthetic SCE when the actual atmospheric transmission and instrumental throughput are measured directly.  We then calculate the synthetic SCE for non-stellar objects, e.g. SNe Ia and galaxies, at different redshifts in Section 5. Section 6 gives a discussion about the possible SCE in ground-based differential photometric transit observations, followed by the conclusions in Section 7.

\section{VARIATION IN THE SYSTEM RESPONSE}\label{sec:sys}
Given a specific flux of an object at the top of the atmosphere, $F_\nu(\lambda)$, the total ADU counts $F$ that are measured by a camera with a photon detector (e.g., Charge Coupled Device, or CCD) can be calculated as:
\begin{equation}\label{eq:photeq}
F = C \int_{0}^{\infty} F_\nu(\lambda)S_b(\lambda)\lambda^{-1}d\lambda
\end{equation} 
Here $S_b(\lambda)$ is the system response function for a given bandpass $b$. The system response includes the Earth's atmospheric transmission along the line-of-sight, the reflectivity of the mirrors on the telescope,  the transmission of the camera lenses and filters, and the quantum efficiency of the detector. $C$ is a constant and related to the effective collecting area of the primary mirror $A$, the inverse gain of the CCD $g$ ($electron/ADU$) and the exposure time $\Delta t$:
\begin{equation}
C \propto \frac{A \Delta t}{g}
\end{equation}

The constant $C$ is not strictly necessary for the calibration, as the observations of spectrophotometric standards (such as DA white dwarfs) using the same instrument can tie a specific natural system onto an AB magnitude system without knowing the actual value of $C$.

\cite{Stubbs2006} proposed that the process of photometric calibration can be separated into the measurement of the atmospheric transmission and a measurement of the instrumental throughput, so that the system response could be separated as
\begin{equation}
S_b(\lambda) = S^{atm}(\lambda) \times S_b^{inst}(\lambda)
\end{equation}

The atmospheric transmission $S^{atm}(\lambda, alt, az, t)$ could change over time and could also depend on the position of the object in the sky $(alt, az)$. It may vary in both a grey-scale (wavelength-independent) and a non grey-scale (wavelength-dependent or the $shape$ of the transmission curve) manner.  Studies have shown that the atmosphere, especially the precipitable water vapor, is homogeneous across the sky~\citep{Querel2014, Li2014}; we therefore do not discuss the spatial variation of atmospheric conditions in the rest of the paper.

The instrumental throughput $S_b^{inst}(\lambda, x, y, t)$ is similar, except that it may vary over time as well as over the position $(x,y)$ on the detector focal plane. Again, the throughput can also vary in both a grey-scale and a non grey-scale manner. It is therefore convenient to separate the system response into a wavelength-independent normalization factor $N$ and a wavelength-dependent $shape$ factor $\phi(\lambda)$ 
\footnote{$\phi(\lambda)$ here is a scaleless function, i.e. only the $shape$ matters. For simplicity, we can define it as $\phi_b(\lambda) = \frac{S_b(\lambda)}{\int S_b(\lambda)d\lambda}$ and $N = \int S_b(\lambda)d\lambda$. This definition is different from~\citet{Ivezic2007}, where they defined $\phi_b(\lambda) = \frac{\lambda^{-1}S_b(\lambda)}{\int \lambda^{-1} S_b(\lambda)d\lambda}$. 
Our definition of $\phi(\lambda)$ represents what we measure from the auxiliary calibration systems, and thus does not include the $\lambda^{-1}$ factor. } 
for each bandpass $b$:
\begin{equation}
S_b(\lambda) = N_{atm} \times \phi^{atm}(\lambda) \times N^b_{inst} \times \phi_b^{inst}(\lambda)
\end{equation}

Equation \ref{eq:photeq} then can be rewritten as 
\begin{equation}
F = C \times N_{atm} \times N^b_{inst} \int_{0}^{\infty} F_\nu(\lambda)\phi^{atm}(\lambda) \phi_b^{inst}(\lambda) \lambda^{-1}d\lambda
\end{equation}

Over a wide area imaging survey that might be conducted for months or years, both $N$ and $\phi(\lambda)$ could be slightly different from one exposure to another, or even within one exposure. For example, airmass extinction and clouds affect $N_{atm}$ ; dust on the mirror affects $N_{inst}$. As described in Section 1, multiple tilings of the survey area, with the repeated measurements of a large number of stars, allow the monitoring of the zeropoint offsets over time and the illumination correction over the focal plane. This paper will not discuss the calibration for the grey-scale variation, i.e. variation of $N$, as the grey-scale correction procedure mentioned in Section \ref{sec:intro} is adequate to calibrate those variations. In this paper we will focus on the variation of the $shape$ of the system response (i.e. atmospheric transmission + instrumental throughput), $\phi_b(\lambda) = \phi^{atm}(\lambda, t) \times \phi_b^{inst}(\lambda, x, y, t)$. The variation of $\phi_b$ will essentially induce the SCE.

\subsection{Variation in the Atmospheric Transmission $\phi^{atm}$}

Atmospheric transmission in the wavelength range covered by DES (300nm--1100nm) is mainly
determined by the following four processes in the Earth's atmosphere \citep{Stubbs2007}: Rayleigh scattering from molecules, aerosol scattering from small particles, molecular absorption, in particular by $O_2$, $O_3$, and $H_2O$, and cloud extinction. The size of water droplets and ice crystals that make up clouds are larger than the wavelength of visible light, and the attenuation by clouds is wavelength independent \citep{Ivezic2007, Burke2014, Li2014}. Cloud extinction is therefore calibrated with gray-scale corrections so we do not consider it in this paper.

The cross sections of  Rayleigh scattering and aerosol scattering both vary smoothly with the wavelength of the incident light. The optical depth of Rayleigh scattering has a wavelength dependence of $\lambda^{-4}$ \citep{Rayleigh1899}. At zenith, it can be simply scaled with the barometric pressure $P_0$ \citep{Hansen1974}. 

The optical depth of aerosol scattering at zenith could be approximately described by the $\textup{\AA}$ngstr\"om formula~\citep{Angstrom1924}:
\begin{equation}\label{eq:aerosol}
\tau_{aerosol} = \tau_0 (\frac{\lambda}{\lambda_0})^{-\alpha}
\end{equation}
where $\tau_0$ is the aerosol optical depth at reference wavelength $\lambda_0$, where $\lambda_0=550$~nm is a convenient reference wavelength.  The $\textup{\AA}$ngstr\"om exponent $\alpha$ is inversely related to the average size of the particles in the aerosol: the smaller the particles, the larger the exponent.  In general, $\alpha$ ranges from 0 to 3 with typical values around 1 to 1.3, depending on the geographic location. $\alpha <$ 1 indicates size distributions dominated by coarse mode aerosol that are usually associated with dust and sea salt, and $\alpha >$ 2 indicates size distributions dominated by fine mode aerosols that are usually associated with urban pollution and biomass burning \citep{Schuster2006}. $\alpha =$ 0 and $\alpha =$ 4 are essentially the two extreme cases of cloud extinction and Rayleigh scattering.

Absorption by molecules only occurs at specific wavelengths. The strong absorption lines by $O_2$ at 690 nm and 760 nm (Fraunhofer ``B" and ``A" bands) are saturated and are closely proportional to the square root of the barometric pressure, so they can be computed and scaled with Rayleigh scattering~\citep{Burke2010}.  $O_3$ absorption mainly affects atmospheric transmission shortwards of 350 nm and in the Chappuis band (450--700 nm).  The optical depth of ozone scales with the ozone column density. Ozone column density is usually measured in Dobson units (DU). Each Dobson unit is equivalent to a thickness of 0.01 mm of ozone at standard temperature and pressure. $H_2O$ absorption mainly influences the atmospheric transmission at wavelengths longer than 600 nm.  The optical depth can be scaled with the precipitable water vapor (PWV) column density in~mm. 

Both the optical depth and the column density mentioned correspond to the vertical path from the observer's location to the top of Earth's atmosphere. For a given atmospheric condition, the atmospheric transmission $T$ of the light also depends on the airmass of the observation. For molecular and aerosol scattering, the transmission $T$ at airmass $X=1$ and at airmass $X=n$ has the simple relation:
\begin{equation}
T(\lambda, X=n) = T^n (\lambda, X=1)
\end{equation}
The transmission due to molecular absorption, however, has a nonlinear curve of growth with respect to the optical depth or the airmass because the absorption departs from the optically thin limit.

We conclude here that the following six parameters determine a specific $shape$ of the atmospheric transmission $\phi^{atm}(\lambda)$: 1) airmass $X$ of the observation, 2) barometric pressure $P_0$ , 3) aerosol optical depth at 550nm  $AOD_{550}$, 4) $\textup{\AA}$ngstr\"om exponent $\alpha$, 5) ozone column density $Ozone$, and 6) precipitable water vapor column density $PWV$. We define the six parameters for a fiducial atmospheric transmission curve at CTIO using $X=1.2$,  $P_0=779$ hpa (1hpa = 100 pascal), Ozone column density $=270$ DU, $PWV = 3$~mm, $AOD_{550}=0.02$, and $\alpha=1$, which are also listed in Table \ref{table:star_error}. We choose an airmass of 1.2 as one that is typical of observations in DES wide-field survey.
Meteorology data from  CTIO show that the average barometric pressure during the year 2014 was 779hpa with standard deviation of 3hpa.  Ozone column density at CTIO ranged from 240 DU to 300 DU with a mean of roughly 270 DU in 2014 according to the NASA Ozone Mapping and Profiler Suite (OMPS) Nadir Mapper.\footnote{http://ozoneaq.gsfc.nasa.gov/tools/ozonemap/} CTIO does not have instrumentation to examine  $PWV$, $AOD$ and $\alpha$. DES therefore deployed the Atmospheric Transmission Monitoring Camera~\citep[aTmCam,][]{Li2012, Li2014} in the summer of 2014, preceded by prototype tests in Oct-Nov 2012 and Sep-Oct 2013, to study the water vapor and aerosol at CTIO. aTmCam is a robotic multi-band imaging system. During DES observations, aTmCam takes simultaneous images in four narrow ($\sim$10~nm) bands centered at 394~nm 520~nm, 854~nm and 940~nm. The aTmCam analysis derives the parameters of the atmospheric transmission models at CTIO including $PWV$ and $AOD$.\footnote{For more information, please see the cited papers or visit http://instrumentation.tamu.edu/aTmCam.html } $PWV = 3$~mm and $AOD_{550}=0.02$ were the average values at CTIO from those early results. The long-term variation range of $PWV$ and $AOD$ at CTIO is not yet clear, but preliminary aTmCam results indicate that that $PWV$ varies between 0--20 mm and $AOD$ varies between 0--0.2. Using the aforementioned parameters, we generated the fiducial atmospheric transmission $\phi^{atm}_{ref}(\lambda)$ using $libRadTran$\footnote{libRadTran is a collection of C and Fortran functions and programs for calculation of solar and thermal radiation in the Earth's atmosphere, see more details at http://www.libradtran.org/} \citep{Mayer2005}, as shown in Figure 1.

\subsection{Variation in the Instrumental Throughput $\phi_b^{inst}$}

\begin{figure}
\centering
\epsscale{1}
\plotone{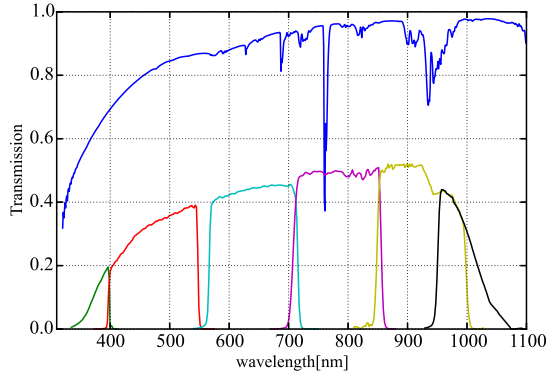}
\caption{The fiducial atmospheric transmission  $\phi^{atm}_{ref}(\lambda)$ at CTIO (upper curve) and the fiducial instrumental throughput $\phi^{inst}_{b,ref}(\lambda)$ of the DES-$ugrizY$ filter bandpasses (lower curves). As DECal only provides a relative throughput measurement, the scale of the lower curves is arbitrary to provide a better visualization.}
\label{fig:std_sys}
\end{figure}

Instrumental throughput is a combination of the mirror reflectivity, lens transmission, filter transmission, and detector sensitivity. The $shape$ of the throughput may vary over time as the detector temperature changes or the filter coatings age. It may also have a spatial dependent variation over the focal plane. Furthermore, each CCD has its own response function. DES has deployed a spectrophotometric calibration system (DECal) that scans the instrument response for all bandpasses by measuring the relative instrumental throughput as a function of wavelength~\citep{Rheault2012}. DECal is used to scan the wavelength range of each filter several times a year, typically during cloudy nights, to monitor the instrumental throughput over time. DECal measurements indicate that the filter bandpass edges vary with focal plane position, primarily with radial position and in $i-$band in particular. This effect is largely due to slightly inhomogeneous filter transmission with incident angle. Here, we define the fiducial instrumental throughput $\phi^{inst}_{ref}(\lambda)$ from the results of DECal scans obtained during Sep-Nov 2013. We use the average throughput over the entire focal plane as the fiducial instrumental throughput, which is also shown in Figure \ref{fig:std_sys}.  

\section{SYNTHETIC SYSTEMATIC CHROMATIC ERRORS}

In this section, we calculate the synthetic SCE when the atmospheric transmission  $\phi^{atm}(\lambda)$ and the instrumental throughput $\phi^{inst}(\lambda)$ deviate from the fiducial values defined in Section 2. We define the synthetic SCE, $\Delta m$, as:

\begin{multline}\label{eq:error}
\Delta m = -2.5 \log_{10} \frac{\int_{0}^{\infty} F_\nu(\lambda)\phi^{atm}(\lambda) \phi_b^{inst}(\lambda) \lambda^{-1}d\lambda}{\int_{0}^{\infty} F_\nu(\lambda)\phi^{atm}_{ref}(\lambda) \phi^{inst}_{b,ref}(\lambda) \lambda^{-1}d\lambda} \\ +   2.5 \log_{10} \frac{\int_{0}^{\infty} F_\nu^{ref}(\lambda)\phi^{atm}(\lambda) \phi_b^{inst}(\lambda) \lambda^{-1}d\lambda}{\int_{0}^{\infty} F_\nu^{ref}(\lambda)\phi^{atm}_{ref}(\lambda) \phi^{inst}_{b,ref}(\lambda) \lambda^{-1}d\lambda} 
\end{multline}

The first term is the change in magnitude for an object with SED $F_\nu(\lambda)$ when $\phi^{atm}$ and/or  $\phi_b^{inst}$ deviate from fidicual values. We use the SEDs of Main-Sequence stars O5V-M6V from the Pickles Atlas (Pickles 1992) as $F_\nu(\lambda)$ for this calculation. The second term is the change in magnitude for a reference star. $\Delta m  = 0$ when the SED of the object is the same as the reference star. This reference star plays a similar role that calibration stars play in the zeropoint computation from the global calibration or the illumination correction in the calibration procedure of DES. The only difference is that the actual survey calibration stars have a range of colors, and the grey-term correction is derived for the average color of the calibration stars.  We pick a solar-type (G2V; $g-i \sim 0.6$) star as the reference star, i.e. a G2V star will have zero $\Delta m$ due to the change of atmospheric transmission and/or instrumental throughput. 

\subsection{Synthetic SCE due to the Variation in Atmospheric Transmission}
We first generate a grid of atmospheric transmission curves for a range of airmass, barometric pressure, AOD, $\alpha$, PWV and Ozone using $libRadTran$.  We then compute SCE due to the variation in atmospheric transmission using Equation \ref{eq:error}  by varying one atmospheric component at a time but keeping the fiducial instrumental throughput unchanged. 

\begin{figure*}[th!]
\centering
\epsscale{1}
\plotone{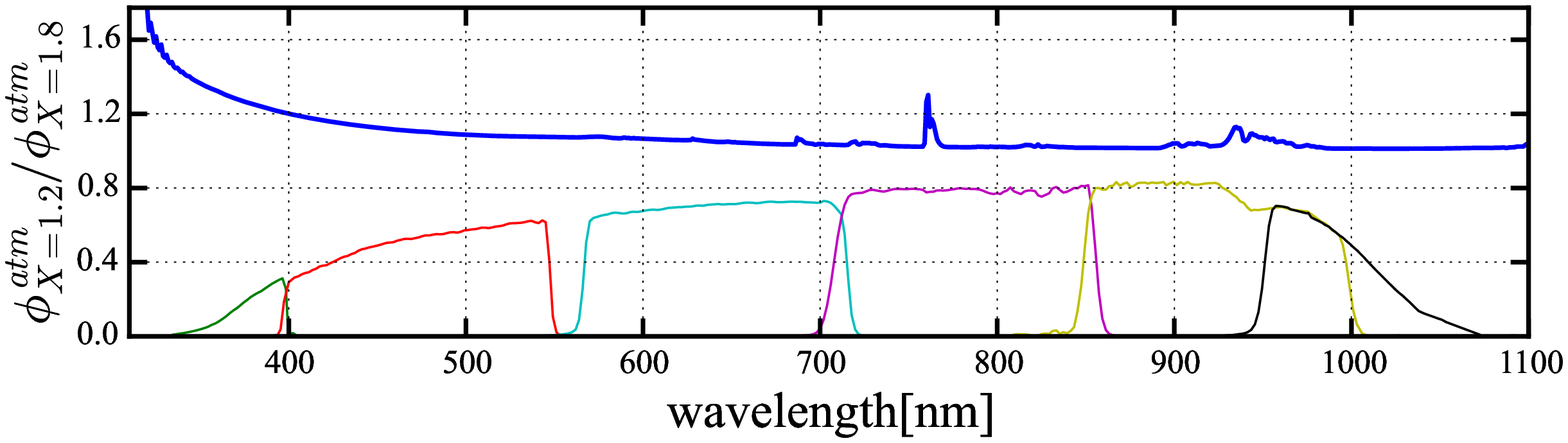}
\plotone{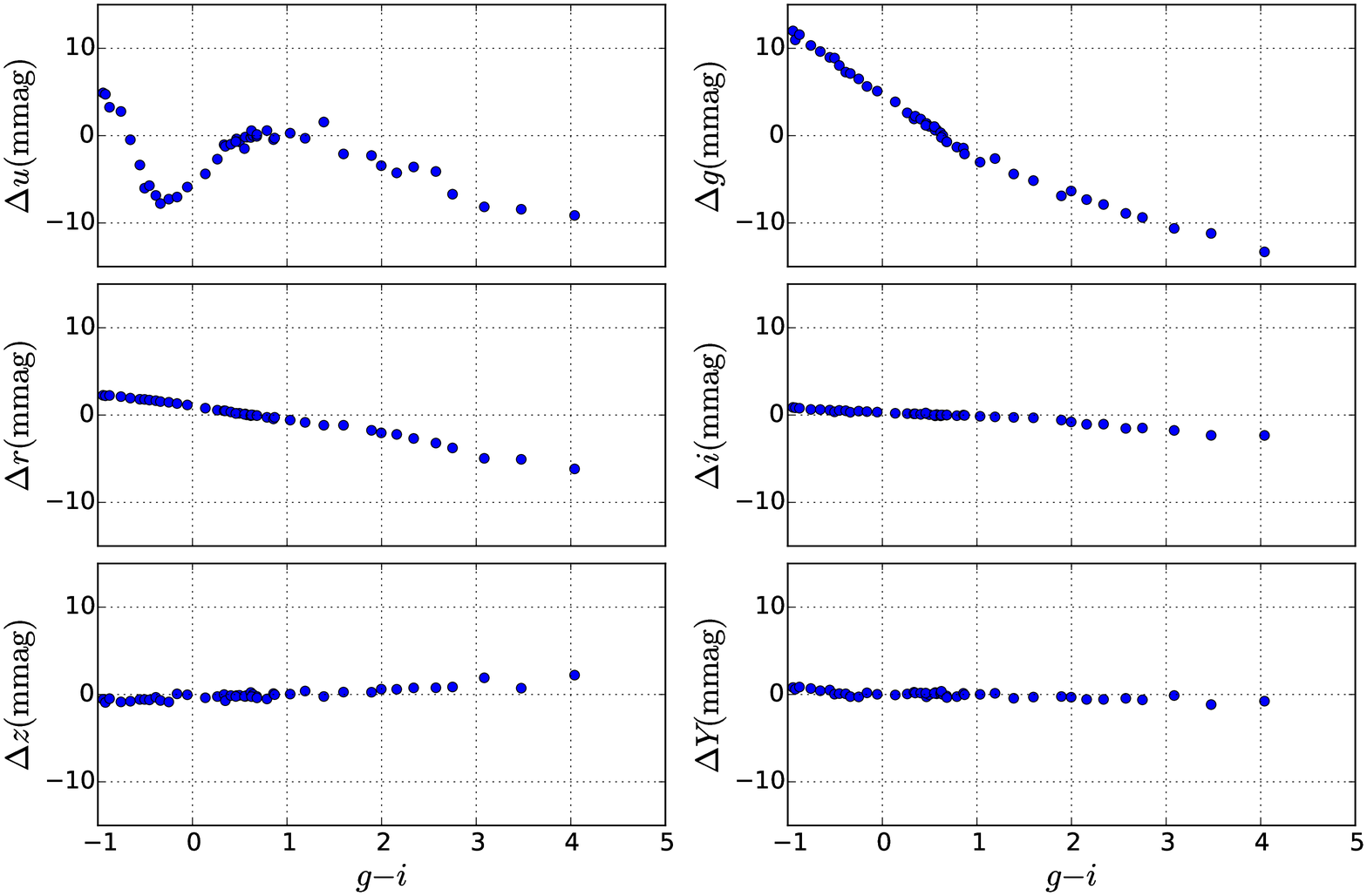}
\caption{Top panel: The ratio of the atmospheric transmission for two airmass values, $X=1.2$ and $X=1.8$. The fiducial instrumental throughput $\phi^{inst}_{ref}(\lambda)$ for $ugrizY$-bands is also shown in the plot as reference. Bottom panels: Synthetic SCE in $ugrizY$-bands for O5V-M6V stars introduced by this airmass change, as a function of $g-i$ color. }
\label{fig:error_airmass}
\end{figure*}

We calculate $\Delta m$ when the airmass changes from $X=1.2$ to $X=1.8$. Figure \ref{fig:error_airmass} shows the ratio of atmospheric transmission at two different airmasses and $\Delta m$ for O5V-M6V stars introduced by this airmass change, as a function of $g-i$ color. $\Delta m$ due to the airmass change is more than $\pm$10mmag in $g$-band for O stars and M stars, and a few mmag in $u$- and $r$-band. $\Delta m$ in $i$-, $z$- and $Y$-band is small. The SCE due to the airmass change are essentially the ``second-order extinction coefficient" or ``airmass color extinction coefficient", which is known to increase towards bluer wavelengths \citep{Henden1990}.

\begin{figure*}[th!]
\centering
\epsscale{1}
\plotone{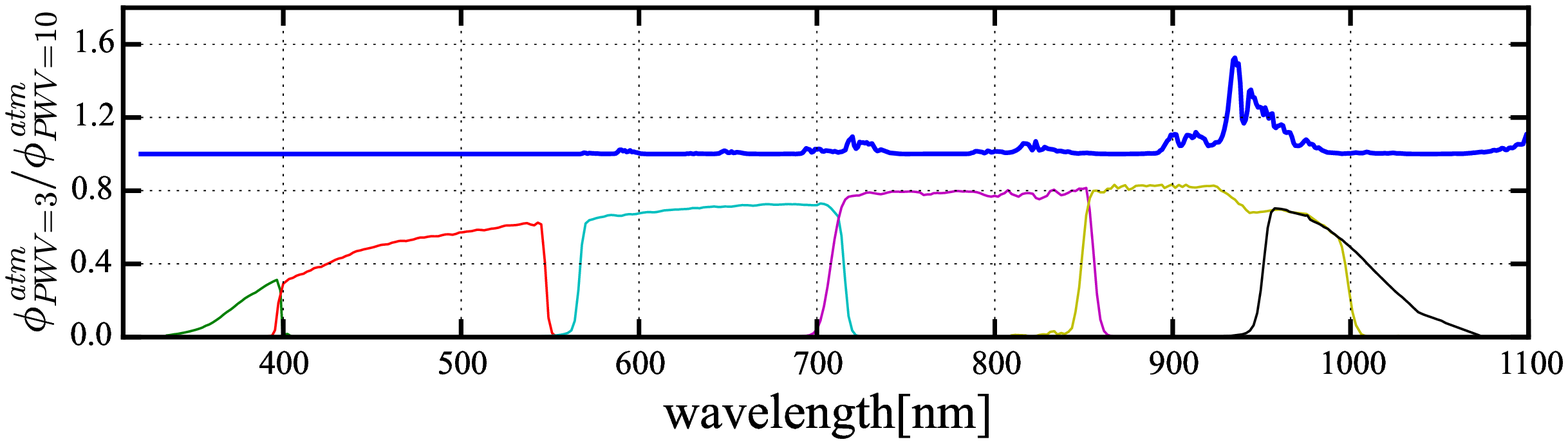}
\plotone{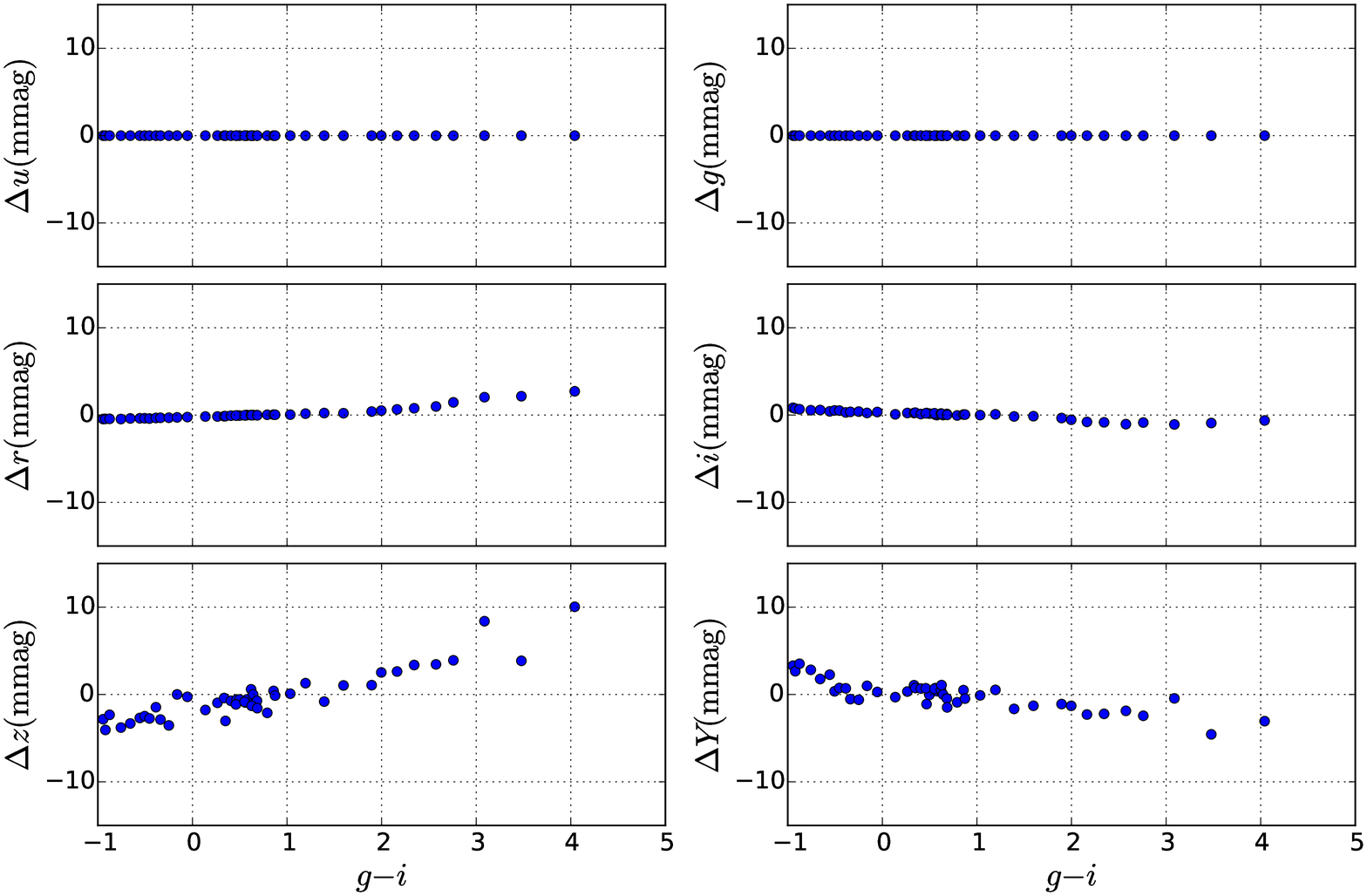}
\caption{Top panel: The ratio of the atmospheric transmission for two PWV values, $PWV=3$ mm and $PWV=10$ mm. Bottom panels: Synthetic SCE in $ugrizY$-bands for O5V-M6V stars introduced by this PWV change, as a function of $g-i$ color.}
\label{fig:error_pwv}
\end{figure*}

We run a similar calculation for the change in PWV from $PWV=3$~mm to $PWV=10$~mm  and show the results in Figure \ref{fig:error_pwv}. $\Delta m$ due to the PWV change is mainly in the $z$- and $Y$-band. The errors can be as large as $+$10 mmag in $z$-band and $-4$ mmag in $Y$-band and thus an error of $>$ 10 mmag in $z-Y$ color. As mentioned earlier, molecular absorption does not vary linearly with column density, and therefore $\Delta m$ caused by a PWV change from  $PWV=3$ mm to $PWV=10$ mm is about the same as that from $PWV=0$ to $PWV=3$ mm or from $PWV=10$ mm to $PWV=20$ mm. 

We also perform a similar calculation for $\Delta m$ when barometric pressure, aerosol and ozone change in the atmosphere. We list $\Delta m$ in Table \ref{table:star_error} for an M6V star ($g-i \sim 4$) as a summary for all above cases, after the grey-term has been removed using a G2V star. $\Delta m$ caused by a change of the barometric pressure is very small. An extreme case of the barometric pressure change from 779 hpa to 789 hpa results in photometric errors of no more than 0.5 mmag in any band.

$\Delta m$ caused by ozone variation is also small; an ozone change from 270 DU to 230 DU, which is an extreme case of the smallest ozone column density measured by the OMPS Nadir Mapper at the longitude and latitude of CTIO, results in $\Delta m < 1$~mmag in any band, and of only a few tenths of mmag in $g$- and $r$-band due to the Chappuis band. The DECam optics essentially has no throughput below 350~nm, so the ozone variation impact on $u$-band photometry is also negligible. This might not be the case for other cameras with greater response below 350~nm.

The change in aerosol optical depth affects mostly $g$- and $r$-band. As shown in Table \ref{table:star_error}, increasing AOD$_{550}$ from 0.02 to 0.20 and keeping $\alpha$ unchanged results in $\Delta m=-11$~mmag in $g$-band. If the increase of the aerosol optical depth is due to the larger size of the aerosol particles, then $\alpha$ would decrease and $\Delta m$ would be smaller compared to the unchanged $\alpha$ case. Equation \ref{eq:aerosol} shows that an increase of AOD makes the atmospheric transmission spectrum redder and a decrease of $\alpha$ makes the transmission spectrum bluer. Therefore, larger particle size (i.e. smaller $\alpha$) with larger AOD might introduce very small $\Delta m$ in one or more bands. AOD and $\alpha$ are somewhat degenerate for the $shape$ of the atmospheric transmission.

The synthetic SCE in Table \ref{table:star_error} are calculated when one of the atmospheric components changes from the fiducial while the other components remain unchanged. Under some conditions, the SCE can be significantly larger. For example, $\Delta m$ for a PWV change from 3 mm to 10 mm will be much larger than 10 mmag if the airmass is at $X=2.0$ instead of the fiducial $X=1.2$. Of course,  cumulative effects can also be larger.

In this section, we calculated the synthetic SCE on stellar photometry caused by the variation in the atmospheric transmission. The SCE caused by the variations in barometric pressure and ozone are very small. Variations in airmass and aerosol mainly affect the DES photometry in the $ugr$-bands; variations in PWV mainly affect the DES photometry in the $zY$-bands, as shown in Table \ref{table:star_error}. This is the primary reason that DES built and deployed aTmCam: to measure the PWV and aerosol at CTIO during DES operations. Furthermore, in order to provide a cross-check of the amount of PWV measured by aTmCam, DES has also installed a high-precision dual-band Global Positioning System (GPS). The GPS is used to measure the PWV, as the variation of PWV affects arrival time of the GPS signal via the increased index of refraction~\citep{Blake2011}. The measured PWV by aTmCam and GPS agrees within the joint uncertainties of the two measurements. More details about a direct comparison can be found in~\citep{Li2014}.

\begin{table*}[th!]
\small
\centering
\caption{Synthetic SCE on a M6V star ($g-i \sim 4$) with respect to a G2V star ($g-i \sim 0.6$).}
\begin{tabular}{c|c|c|cccccc}
\hline
\multirow{2}{*}{Component} & \multirow{2}{*}{Fiducial} & \multirow{2}{*}{Changed$^a$} & \multicolumn{6}{c}{Synthetic SCE (mmag)} \\ \cline{4-9} 
                      &                     &                     & u  & g  & r  & i  & z  & Y \\ \hline
Pressure & $P_0=779$ hpa    & $P_0=789$ hpa & $-$0.2 & $-$0.4 & $-$0.08 & $-$0.06 & 0.04 & $-$0.02 \\ \hline
\multirow{2}{*}{Aerosol}     & AOD$_{550}=0.02$, $\alpha=1$ & AOD$_{550}=0.20$, $\alpha=1$  & $-$3 & $-$11 & $-$6 & $-$4 & $-$1 & $-$0.2 \\ \cline{2-9} 
          & AOD$_{550}=0.02$, $\alpha=1$ & AOD$_{550}=0.20$, $\alpha=0.5$  & $-$41 & $-$5 & $-$3 & $-$2 & $-$0.8 & $-$0.1 \\ \hline
 PWV       & PWV $=3$ mm    & PWV $=10$ mm &  0 &  0 & 3 & $-$0.6 & +10 & $-$3\\ \hline

 Ozone     & Ozone $=270$ DU & Ozone $=230$ DU & 0 & $-$0.7 & 0.9 & 0.1 & 0 & 0 \\ \hline
 Airmass   & $X=1.2$      & $X=1.8$ & $-$9 & $-$13 & $-$6 & $-$2 & +2 & $-$0.8\\ \hline
  Instrument & DECal scan   & shift 2 nm & $-$24 & $-$15 & $-$16 & $-$19 & $-$10 & $-$5 \\ \hline

\end{tabular}
    \begin{tablenotes}
       \item $^a$The ``changed" conditions here are just examples. For pressure and ozone, we used the extreme examples since the SCE are small; for aerosol, PWV and airmass, we give the examples where the change could introduce about 1\% or 10~mmag SCE. For the instrument, we choose 2nm shift as it's about the average value from the DECal scans.
    \end{tablenotes}
   
\label{table:star_error}
\end{table*}

\subsection{Synthetic SCE due to the Variation in Instrumental Throughput}
In this section, we study the synthetic SCE, $\Delta m$, due to the variation in the instrumental throughput. For reference, SDSS discovered variations of the instrumental throughput over its survey period \citep{Doi2010}. Such variations could introduce SCE similar to those caused by the variation in the atmospheric transmission.  DES so far has not seen a variation in instrumental throughput over time from the DECal scans in the past three years; however, data from DECal have shown a shift of either the blue or red edges of the filter bandpasses over the focal plane. 

\begin{figure*}[th!]
\centering
\epsscale{1}
\plotone{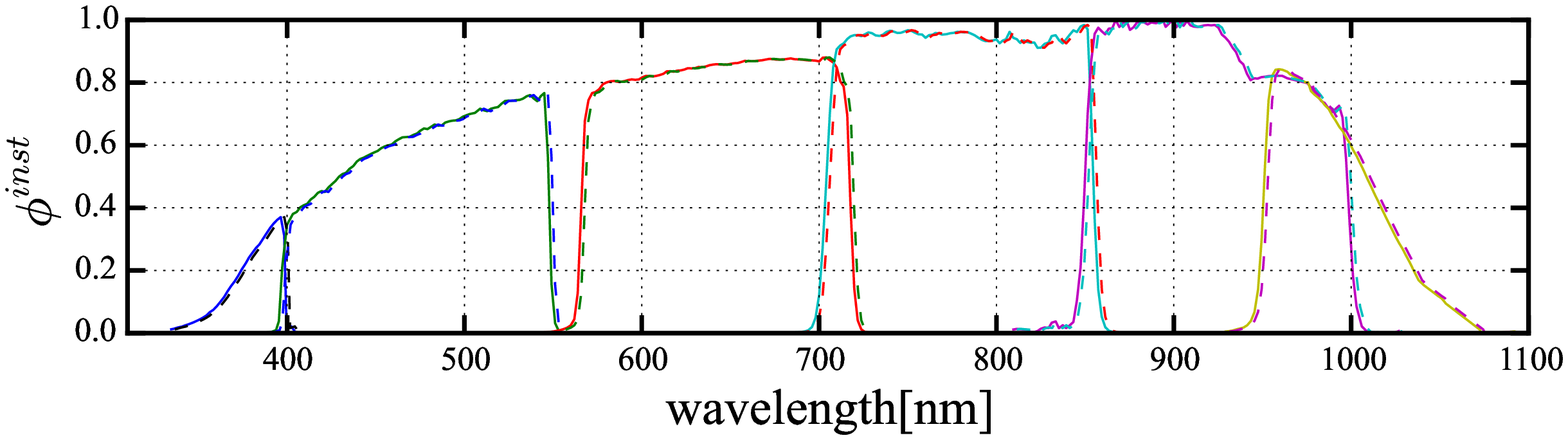}
\plotone{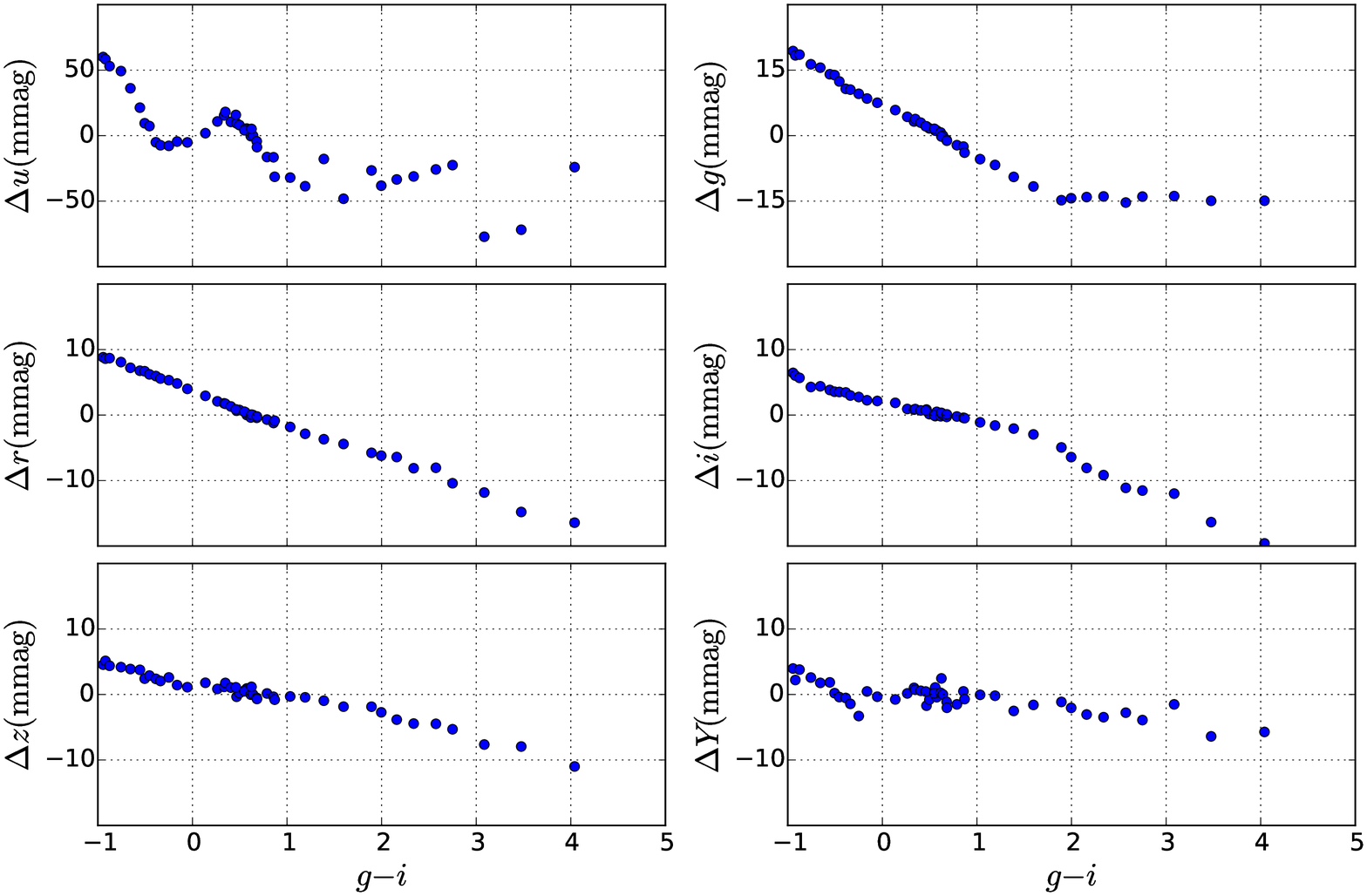}
\caption{Top panel: Solid lines are the fiducial instrumental throughput $\phi^{inst}_{ref}(\lambda)$ for $ugrizY$-bands.  Dashed lines are the instrumental throughput with a 2 nm shift towards the longer wavelength. Bottom panels: Synthetic SCE in the $ugrizY$-bands for O5V-M6V stars introduced by this 2 nm  bandpass shift, as a function of $g-i$ color. Note that the scales for the $u-$ and $g-$band are different from the other four bands.}
\label{fig:error_bandpass_shift}
\end{figure*}

We shift the fiducial instrumental throughput $\phi^{inst}_{ref}(\lambda)$ 2 nm towards the longer wavelength and define it as a changed instrumental throughput $\phi^{inst}(\lambda)$. We then calculate $\Delta m$ due to this shift for O5V-M6V stars, shown in Figure \ref{fig:error_bandpass_shift} and Table \ref{table:star_error}. We again use a G2V star as reference to remove the grey term, as this is removed by the illumination correction using star-flats in the calibration procedure of DES. Except for $u$-band, $\Delta m$ in the other 5 bands are at the level of 1-2\%. The actual bandpass shifts from the DECal scans in $griz$ bandpasses are roughly 1, 3, 6 and 2 nm respectively, but only one of the bandpass edges shift, instead of both. The $u$- and $Y$-band show almost no edge shift. More details about DECal and bandpass variations will be presented in Marshall~et~al.~in prep.

\section{SYSTEMATIC CHROMATIC ERRORS IN STARS AS SEEN IN DES OBSERVATIONS}

As noted earlier, DES obtains its data with the DECam camera on the Blanco 4m telescope at CTIO. The data are transferred in near real-time over the course of each night to the  National Center for Supercomputing Applications (NCSA) at the University of Illinois at Urbana-Champaign, where the DES Data Management (DESDM) team performs an initial nightly processing of the data, including image detrending, cataloging, and astrometric calibration of the individual exposures.  There is also an annual re-processing, which includes a full re-processing of the single-epoch exposures, a global photometric calibration of the data, and a coadd of overlapping exposures.  An overview of the process is described in \citet{Balbinot2015}, and details can be found in Gruendl et al. (in preparation). The current global photometric calibration only considers the grey-scale zeropoint correction, and therefore the SCE have not been corrected in the current catalog. In order to show that SCE exist in the catalog, we used the calibrated photometry derived using single epoch exposures from DES.

In this section, we give two examples using a test sample from DES data. We present the SCE due to the PWV variation and due to the location on the DECam the focal plane. We show that SCE in DES observations match synthetic SCE to within a few mmag, which suggests that corrections based on measurements from the auxiliary calibration system (aTmCam + DECal) can be used to significantly improve photometric precision.

\subsection{SCE due to PWV Variation}
We first show how the change of the PWV in the atmosphere affects the photometry in DES. The PWV was measured by aTmCam during DES observations.

\begin{figure*}[th!]
\centering
\epsscale{1}
\plottwo{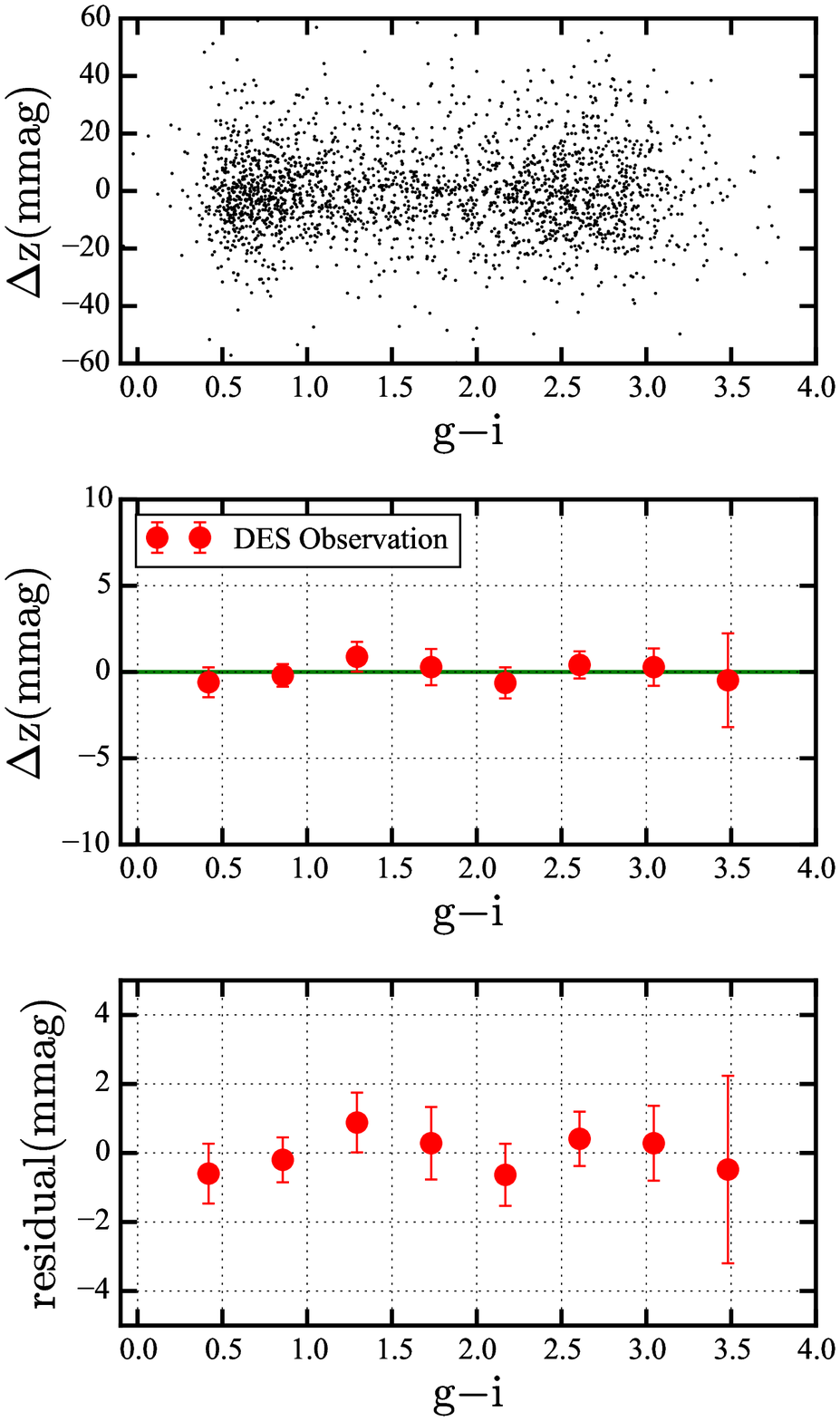}{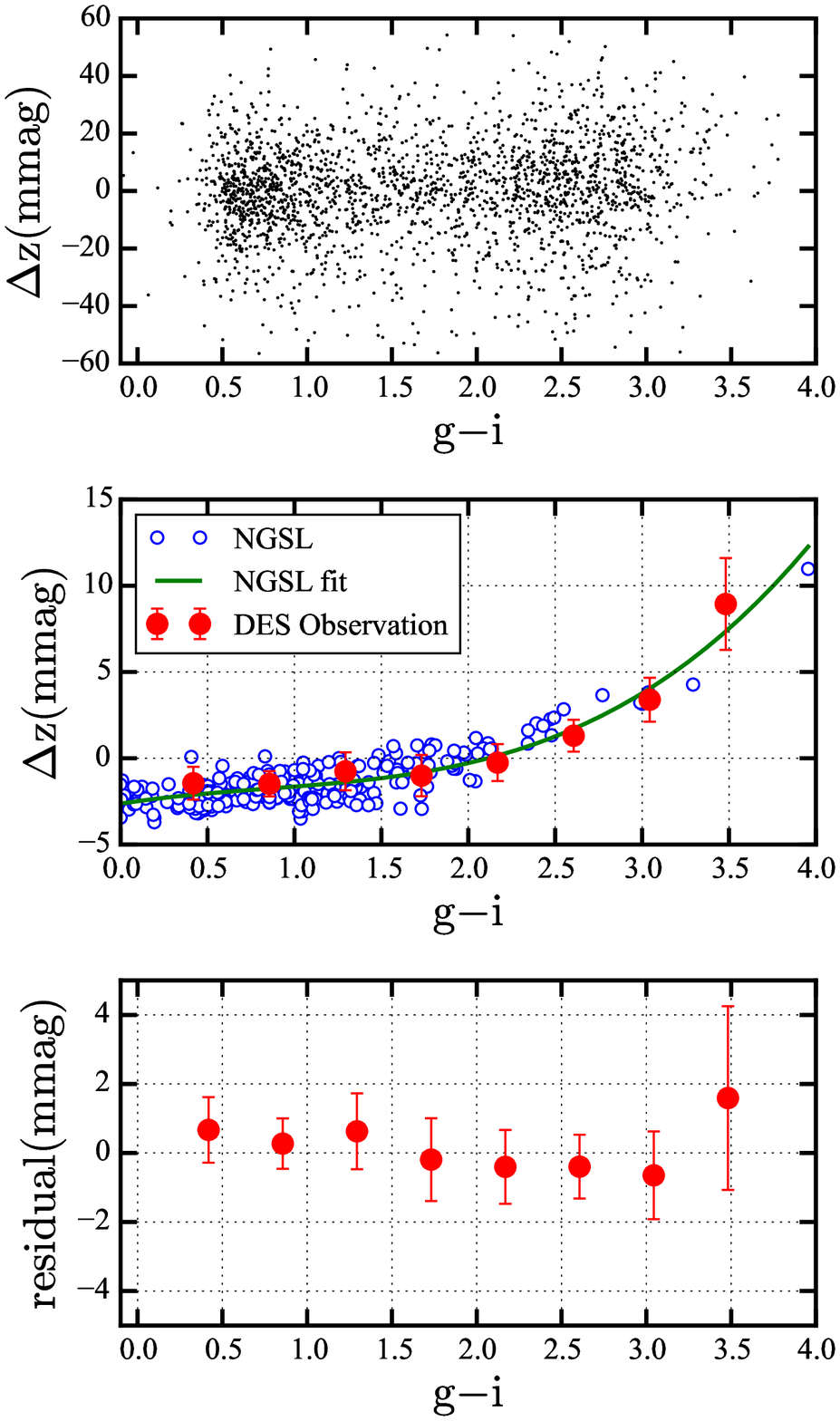}
\caption{Measurements of the differences of $z$-band photometry, $\Delta z$, on two nights with similar PWV (left) and two nights with different PWV (right), as a function of the $g-i$ color of stars. Top panels: Each black dot is a $\Delta z$ from one star. Middle panels: Stars with $0.2 < g-i < 3.7$ are divided into 8 bins and the average of $\Delta z$ in each bin is calculated and shown as the red filled circles. The error bars show the error of the mean in each bin.  A green line of $\Delta z=0$ is shown on the left as the PWV is similar on those two nights. On the right panel, the green line is a third-order polynomial fit to the synthetic SCE calculated using the atmospheric transmission models and the stellar templates, shown as the blue open circles. Bottom panels: Residuals of the average of $\Delta z$ in each bin, after corrections with the fit to the synthetic errors (i.e. green lines) in the middle panels.}
\label{fig:pwv}
\end{figure*}

Standard star fields are observed on every photometric night in the DES during evening and/or morning twilight.  \ 
For one of DES standard star fields SDSSJ2300+0000 ($\sim$3~deg$^2$), Figure \ref{fig:pwv} shows the $z-$band stellar photometry difference, $\Delta z$, between two exposures taken on different nights, as a function of $g-i$ color of the stars. $\Delta z$ is derived from the photometry in the single epoch catalog from the standard star calibration exposures.  The position of the stars on the CCD focal plane in two standard star exposures is constant to within the pointing error of the telescope (roughly 5 arcsec), so there are no significant SCE from the variation of the instrumental throughput over the focal plane. \
The exposures are selected to be taken at the same airmass to ensure there are no SCE from the airmass change. The selection criteria of -0.002 $<$ $spread\_model\_i$\footnote{Spread model is a parameter measured by SExtractor~\citep{Bertin1996}. It describes whether an object is better fit by the PSF or a broadened version of the PSF. It may be used as an indicator for star-galaxy separation.\citep{Desai2012}} $<$ 0.002 and $z < 18$~mag are applied in order to ensure that the targets are all bright stars so that the statistical errors from photon fluctuations are negligible ($<$ 5mmag on average). \
We adopted the $g-i$ color for each star from the photometry in the coadd catalog. Since the coadd photometry is essentially the average over many exposures taken under different conditions, the $g-i$ color from coadd catalog is averaged over different SCE and might be slightly different from the $g-i$ color from a single-epoch. However, this should not be a problem as such a color difference would be a second-order effect to the SCE and should be negligible (i.e. $<1~$mmag).

We first calculate $\Delta z$ for two exposures for nights 2014-11-13 and 2014-11-12, which has $PWV=3.6$~mm and $PWV=4.2$~mm from the measurements by aTmCam, shown in the top left panel of Figure \ref{fig:pwv}; we then perform the same calculation on nights 2014-11-13 and 2014-11-15, which have $PWV=3.6$~mm and $PWV=13.6$~mm, shown on the top right panel. In both cases, there are more than 3000 stars matched from 2 exposures. There is an obvious trend in the top right panel showing that $\Delta z$ is correlated with $g-i$ when there is a large difference between the PWV values. \

  We divide these stars into 8 equal-width bins over the range $0.2 < g-i < 3.7$. Except for the last bin which only has about 60 stars, all bins have more than 300 stars.\
  We calculate the average of $\Delta z$ in each bin, shown as the red filled circles in the middle panels of Figure \ref{fig:pwv}. The error bars show the error of the mean in each bin. On nights with similar PWV, the average of $\Delta z$ is  consistent with zero for all types of stars, as shown in the middle left panel. \
  However, on the nights with a large difference in PWV, the average of $\Delta z$ deviates from zero for red stars, as shown in the middle right panel. The most significant difference is at $g-i\sim3.5$, where $\Delta z$ is almost 4-$\sigma$ away from zero. This is strong evidence showing the existence of SCE when the PWV changes.  
  
  We calculated the synthetic $\Delta z$ when the atmospheric transmission changes from PWV=3mm to PWV=13mm. We followed the same steps as discussed in Section 3.1, except that instead of using a G2V star as the reference star, we used stars with $g-i \sim 2$ as reference stars to remove the grey-term variation, as stars with $g-i \sim 2$ tend to have zero errors in $\Delta z$ between these two exposures. Furthermore, instead of using the Pickles Atlas, we used the stellar template from Next Generation Spectral Library\footnote{https://archive.stsci.edu/prepds/stisngsl/} (NGSL), which contains flux calibrated stellar templates for more than 350 stars. We note that the synthetic SCE calculated using NGSL and the Pickles Atlas generally show the same trend. As shown in Figure \ref{fig:error_pwv}, the synthetic SCE using the Pickles Atlas have a few millimag of scatter in $z$- and $Y$-band. We therefore adopted NGSL to compare with DES data as it contains a much larger sample of stellar templates. The synthetic SCE calculated using each stellar template in NGSL are shown as open circles in the middle right panel of Figure \ref{fig:pwv}.
  
  We then fit a third-order polynomial to the computed synthetic $\Delta z$, and show this curve as a green line in middle right panel of Figure \ref{fig:pwv}. The difference between the DES data and a fit to the synthetic SCE using the NGSL templates are shown in the lower panels of Figure \ref{fig:pwv}. The difference is less than 2 mmag over the $g-i$ color range. We emphasize that the fit is $NOT$ to DES data, but rather it demonstrates that the SCE due to the measured PWV change between the two nights are removed exceptionally well.

\subsection{SCE due to Location on the DECam Focal Plane}
Here we show how the variation of instrumental throughput over the DECam focal plane affects the photometry in DES. We use $i$-band as the example since $i$-band has the largest bandpass edge shift measured from the results of DECal scans.

\begin{figure}[th!]
\centering
\epsscale{1}
\includegraphics[scale=0.4]{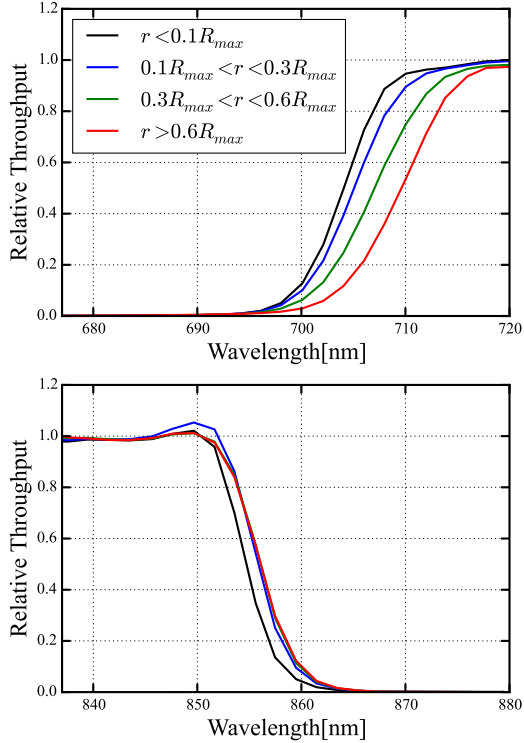}
\caption{
The relative throughput from a DECal scan around the cut-on (cut-off) wavelength in the top (bottom) panel from 4 Regions defined in the text. The throughput is normalized at 770 nm. Note there is about a 6 nm shift from the center of the focal plane to the edge of the focal plane.}
\label{fig:pos}
\end{figure}

We divide the DECam focal plane into 4 regions based on the location of the center of the 62 CCD chips on the focal plane: Region 1 (0-0.1$R_{max}$),  Region 2 (0.1$R_{max}$-0.3$R_{max}$), Region 3 (0.3$R_{max}$-0.6$R_{max}$), and Region 4 (0.6$R_{max}$-$R_{max}$), where $R_{max}$ is the maximum radius of the focal plane.  The relative throughputs from DECal scans around the blue (red) edge are shown in the top (bottom) left panel of Figure \ref{fig:pos}. The average throughput in four different regions is normalized to the value at 770 nm. The figure shows that there is about a 6 nm shift at the blue edge when comparing the center of the focal plane to the edge of the focal plane. 

\begin{figure*}[th!]
\centering
\epsscale{1}
\includegraphics[scale=0.33]{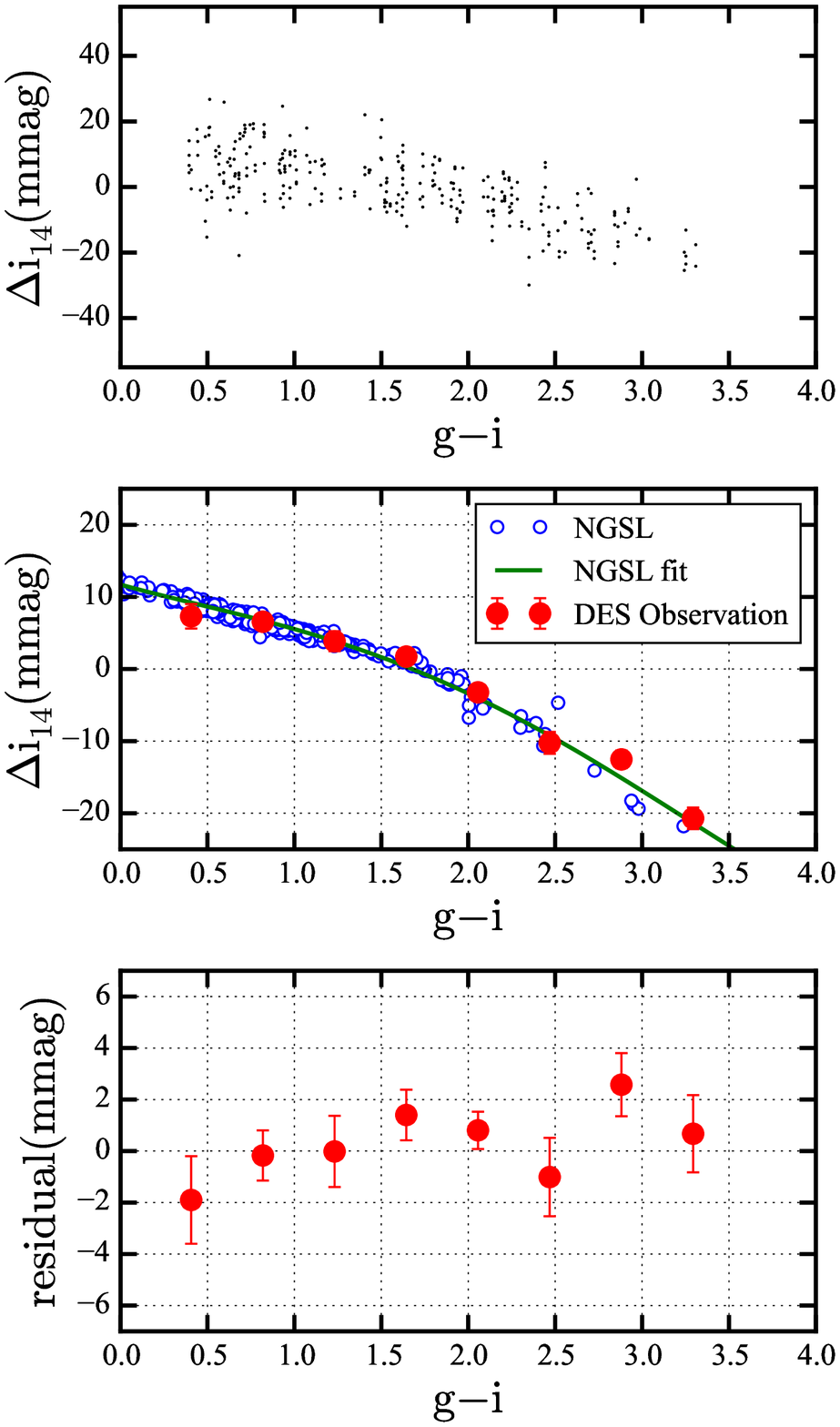}
\includegraphics[scale=0.33]{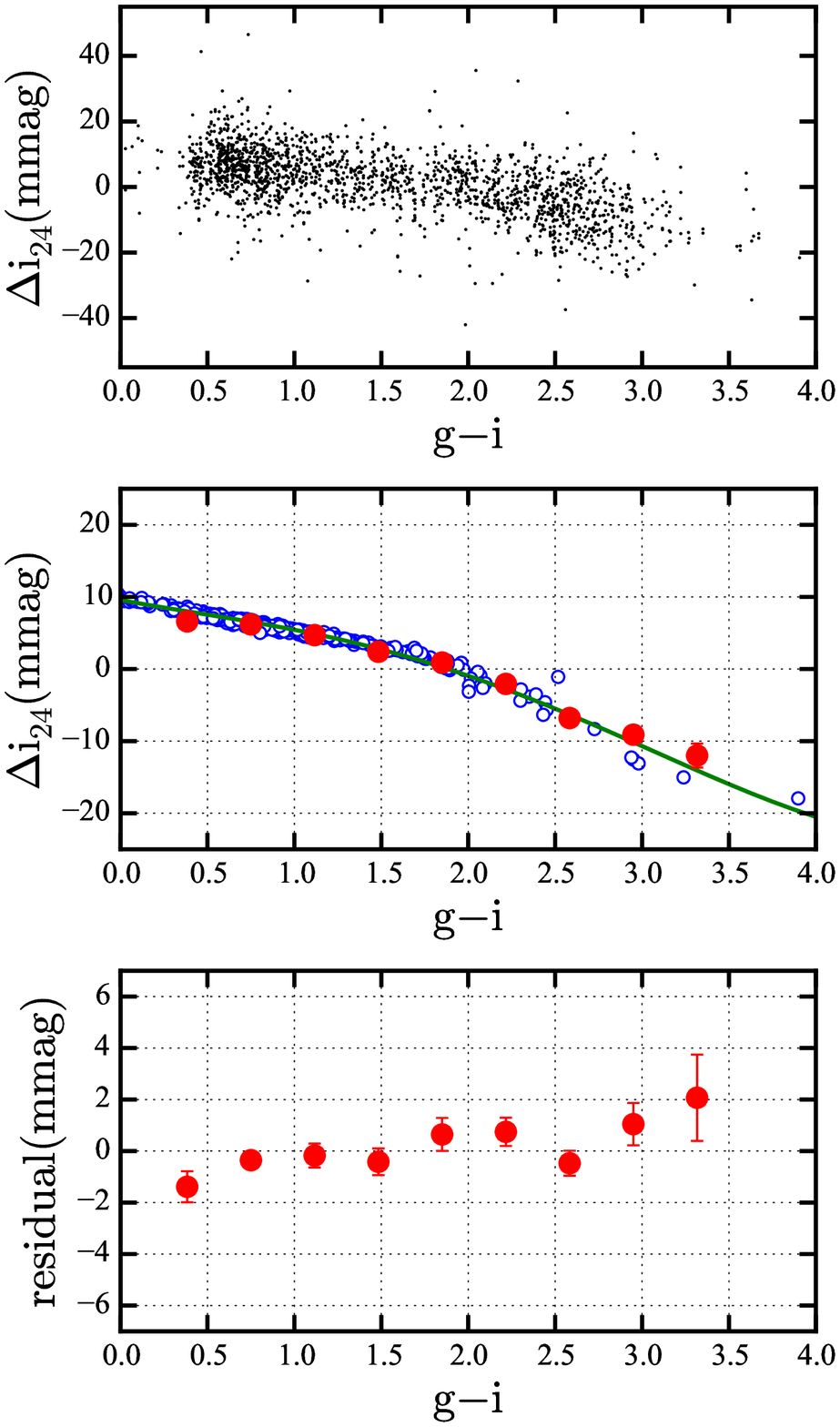}
\includegraphics[scale=0.33]{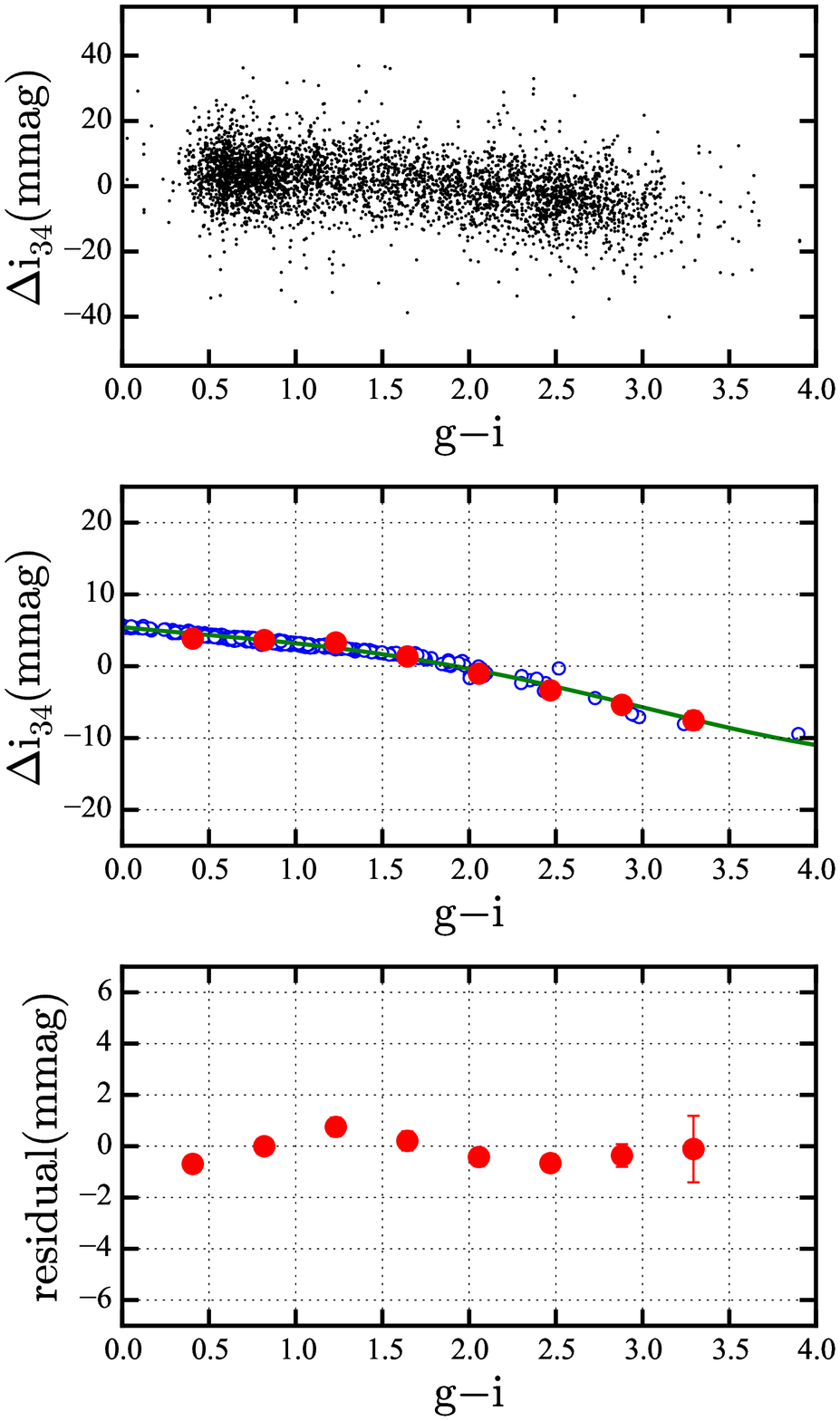}
\caption{
Top panels: $\Delta i$ for Regions 1 \& 4 (left), Regions 2 \& 4 (middle) and Regions 3 \& 4 (right), as a function of $g-i$ color. There are 291, 1890, and 4189 points in the left, middle and right panels, respectively. Middle panels: We divided these stars into 8 bins and then calculated the average of $\Delta i$ in each bin, shown as the red filled circles.  The error bars show the error of the mean in each bin.  Also calculated are the synthetic SCE using the stellar spectra from NGSL, shown as the blue open circles. A fourth-order polynomial fit to the synthetic SCE is shown as the green line. Bottom panels: Residual plots of the binned $\Delta i$ minus the fit from the synthetic SCE. Note that the vertical scale for each panel is different.}
\label{fig:error_pos}
\end{figure*}

We use the calibrated DES data in the same area as the standard star field SDSSJ2300+0000 to calculate the SCE in $i$-band. We use the single epoch results from the survey exposures instead of the standard star calibration exposures. Since the tilings from the survey exposures have some small spatial offset (i.e. ``dithered"), each star has been observed multiple times using different regions of the DECam focal plane. The same selection criteria as in Section 4.1 ($-$0.002 $<$ $spread\_model$ $<$ 0.002 and z $<$ 18 mag) are applied in order to ensure that the targets are all bright stars. We also make an airmass requirement (1.12 $< X <$ 1.22) to ensure minimal SCE from the change of airmass. We then determine the $i-$band magnitude in four regions as described above, $i_1$, $i_2$, $i_3$, and $i_4$ for Region 1 to 4, respectively. For each star observed in Region 4, we find the same star in the other three regions when it is available and calculate the difference $\Delta i$. We found 291 matches for Regions 1 \& 4; the differences $\Delta i_{14}=i_1-i_4$ are calculated and shown in the top left panel of Figure \ref{fig:error_pos}. We bin the DES data in a similar way as in Section 4.1 and calculate the synthetic SCE using the stellar spectra from NGSL, the instrumental throughputs for Regions 1 \& 4 from the DECal scans and the fiducial atmospheric transmission model, shown in the middle left panel of Figure \ref{fig:error_pos}.  A fourth-order polynomial fit to the synthetic SCE is shown as the green line in the same panel. We repeat the calculation and show the $\Delta i$ for Regions 2 \& 4 and Regions 3 \& 4 in the middle column and right column of Figure \ref{fig:error_pos}. There are 1890 matches for Regions 2 \& 4 and 4189 matches for Regions 3 \& 4, respectively. The bottom panels show the residuals of the binned $\Delta i$ after correction using the fit from the synthetic SCE calculation. We show here that $\Delta i$ after the correction is less than $<$ 3mmag for any stars at any position of the focal plane.

The SCE due to position change are essentially the instrumental color-term over the DECam focal plane. It is usually calculated empirically using stars to get the correction to first order. Such empirical linear color corrections are good for stars with $0<g-i<2$. However, for very red stars and non-stellar objects, an empirical linear correction is not sufficient.

\subsection{Residual Errors After Correction}

Above we show that measurements of the atmospheric transmission and instrumental throughput can be used to correct imaging data to high photometric precision. Any system that determines the shape of the atmospheric transmission, however, will not produce perfect results and there will be errors in the determined values of the PWV, AOD, etc. Table \ref{table:star_residual_error} shows the effects of uncertainties in the determination of the important parameters. Notably, measurements of the PWV that are accurate to $\sim$10\% generally are adequate to ensure that residual errors after correction are less than 1 mmag in all DECam bands. Similarly, AOD and $\alpha$ determinations accurate to 0.02 and 0.1 also generally produce corrections that give less than $\sim$1 mmag residual errors.

Measurement of the variation in instrumental throughput across the focal plane generally require determination of the wavelengths of any shift in the bandpass to $\sim$0.2 nm precision. Determination of the bandpass to this level will give $<2$~mmag residual error in most bands. 

\begin{table*}[th!]
\small
\centering
\caption{Residual Errors on a M6V star ($g-i \sim 4$) when a measurement by an auxiliary instrument has small uncertainties.}
\begin{tabular}{c|c|c|cccccc}
\hline
\multirow{2}{*}{Component} & \multirow{2}{*}{Measured} & \multirow{2}{*}{True} & \multicolumn{6}{c}{Synthetic Residual SCE (mmag)} \\ \cline{4-9} 
                      &                     &                     & u  & g  & r  & i  & z  & Y \\ \hline
\multirow{2}{*}{Aerosol}     & AOD$_{550}=0.02$, $\alpha=1$ & AOD$_{550}=0.04$, $\alpha=1$  & $-$0.36 & $-$1.3 & $-$0.68 & $-$0.41 & $-$0.15 & $-$0.02 \\ \cline{2-9} 
          & AOD$_{550}=0.20$, $\alpha=1$ & AOD$_{550}=0.20$, $\alpha=0.9$  & 0.47 & 1.4 & 0.58 & 0.28 &  0.08 &0.01 \\ \hline
 PWV       & PWV $=3$ mm    & PWV $=3.3$ mm &  0 &  0 & 0.25 & $-$0.05 & 0.92 & $-$0.27\\ \hline

  Instrument & DECal scan   & shift 0.2 nm & $-$6.9 & $-$1.5 & $-$1.6 & $-$1.9 & $-$1.0 & $-$0.74 \\ \hline
\end{tabular}

\label{table:star_residual_error}
\end{table*}

\section{SYNTHETIC SYSTEMATIC CHROMATIC ERRORS ON NON-STELLAR OBJECTS}\label{sec:nonstar}

Since the SED of a non-stellar object is significantly different from a star, the SCE on some of the non-stellar objects can be larger than what we have seen for stars. We show two examples in this section: Type Ia supernovae (SNe Ia) and elliptical galaxies. We calculate the synthetic SCE $\Delta m$ for these two types of objects using Equation \ref{eq:error}. As the shape of the SEDs changes with redshift, the $\Delta m$ are also redshift dependent.

\subsection{Type Ia Supernovae}
For the DES survey, the sub-percent photometry precision goal comes from supernova cosmology, which needs precise photometry so that one can measure the luminosity distances of SNe Ia over a wide redshift range. 

\begin{figure*}[th!]
\centering
\epsscale{1}
\plotone{ratio_tran_pwv.eps}
\plotone{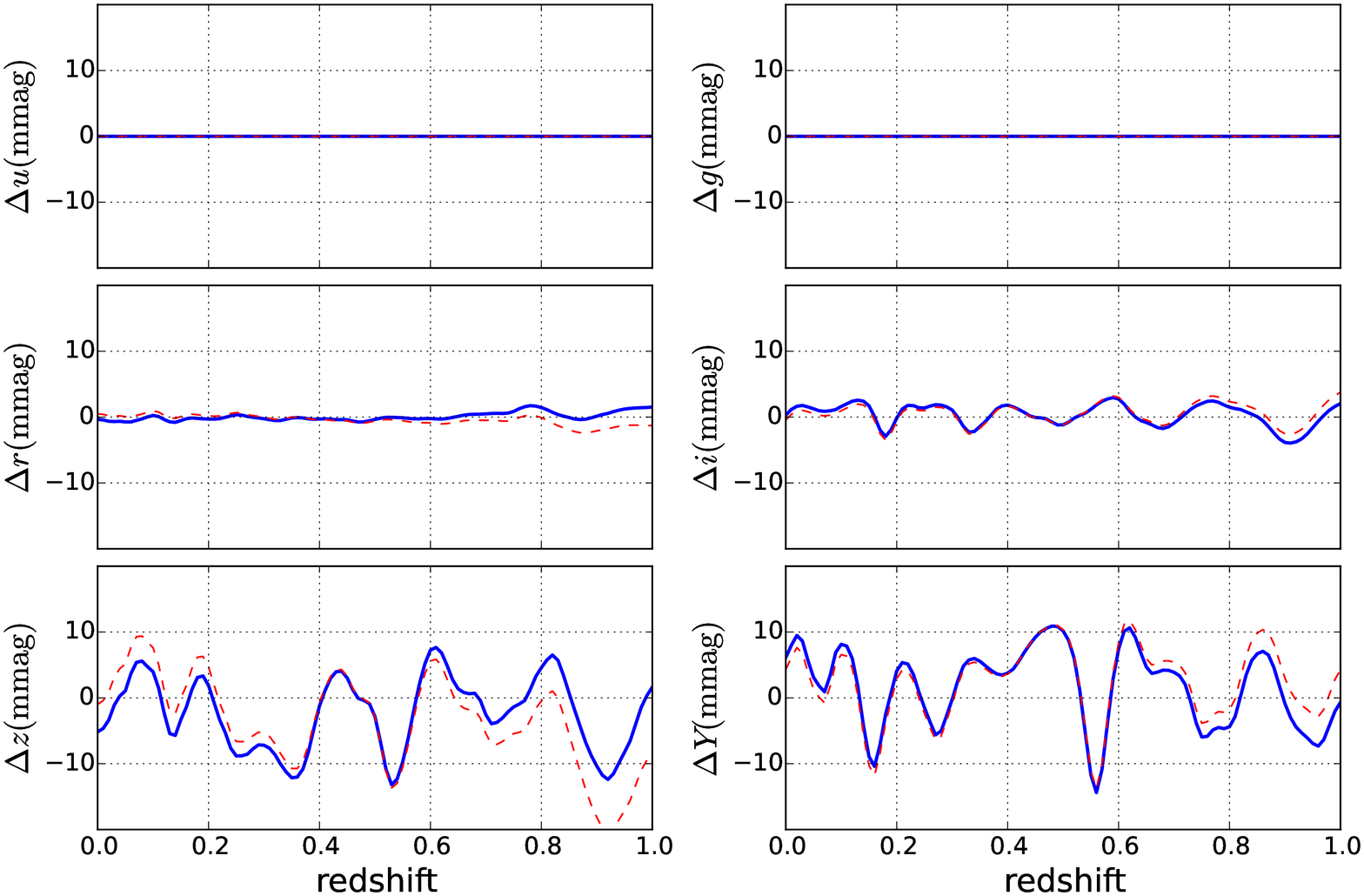}
\caption{Synthetic SCE in $ugrizY$-bands for SNe Ia as a function of redshift when the PWV in the atmosphere changes from 3mm to 10mm, shown as the blue sold lines. The red dashed lines are the residual errors after using the correction derived for stars with the same $g-i$ color as the SNe Ia.}
\label{fig:error_pwv_sn}
\end{figure*}

Here we give an example of how the synthetic SCE change with redshift. Figure \ref{fig:error_pwv_sn} shows $\Delta m$ caused by a PWV change from 3 mm to 10 mm for a Type Ia supernova (SN Ia) as a function of redshift $z$. We use the SED from SN2011fe \citep{Pereira2013} as the template for the calculation. The template was taken 0.27 days prior to maximum brightness. The peak-to-valley errors can be as large as 20 mmag for $z$- and $Y$- band over redshift $z=0$ to $z=1$, shown as the blue solid lines. Because a SN Ia SED is very different from that of a star, the SCE for a star and a SN Ia are very different, even if they share the same $g-i$ color. The red dashed lines in Figure~\ref{fig:error_pwv_sn} show the SCE residuals after naively using the corrections derived from stars with the same $g-i$ color as the SNe Ia, i.e. synthetic SCE calculated in Section 3.1. As the figure shows, the SCE are not properly corrected, and sometimes are even larger.

In Table \ref{table:sn_error}, we summarize the synthetic SCE on SNe Ia by the changes of other atmospheric components or by a 2 nm shift. It is similar to Table \ref{table:star_error} except that the SCE here are calculated as the peak-to-valley SCE over redshift $z=0$ to $z=1$.


\subsection{Galaxies}
\begin{figure*}[th!]
\centering
\epsscale{1}
\plotone{bandpass_shift.eps}
\plotone{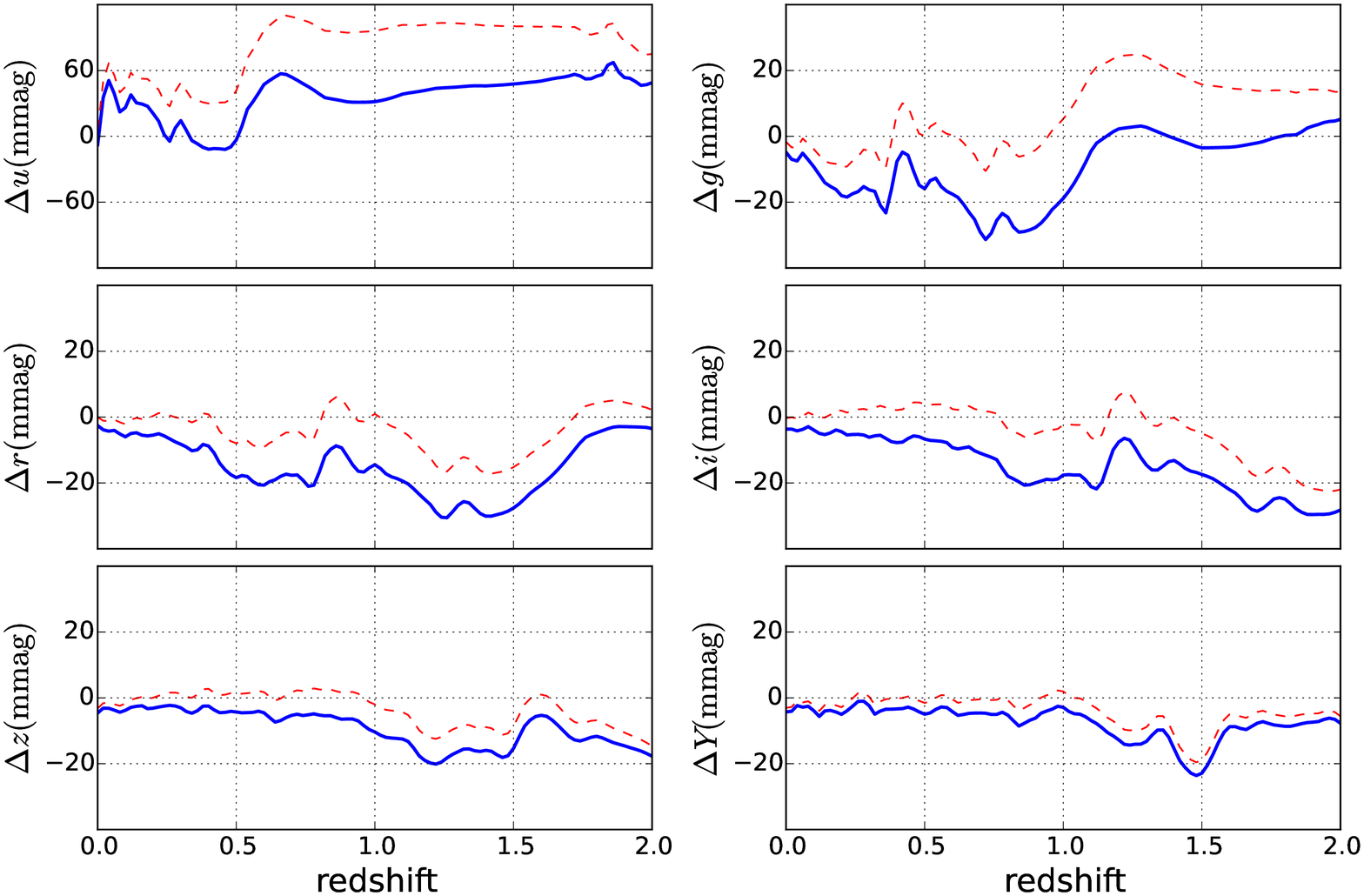}
\caption{Synthetic SCE in $ugrizY$-bands for elliptical galaxies as a function of redshift, when the instrumental throughput is moved by 2 nm toward longer wavelength. The red dashed lines show the residual errors after using the correction derived for stars with the same $g-i$ color as the elliptical galaxies. Note that the scale for $u-$band is different from the other five bands.}
\label{fig:error_bandpass_shift_elgal}
\end{figure*}

Precise photometry helps the determination of photometric redshifts of galaxies~\citep{Wolf2001, Ilbert2006}. We therefore study the synthetic SCE for galaxies. We used a spectral template of elliptical galaxies from \cite{Coleman1980} for this calculation. We summarize the peak-to-valley SCE on elliptical galaxies over redshift $z=0$ to $z=2$ in Table \ref{table:elgal_error}. As an example, Figure \ref{fig:error_bandpass_shift_elgal} shows $\Delta m$ caused by a 2 nm shift of the instrumental throughput towards longer wavelength, as a function of redshift. The drop of $\Delta r$ around redshift $z=0.4$, $\Delta i$ around redshift $z=0.8$, $\Delta z$ around redshift $z=1.1$ and $\Delta Y$ around redshift $z=1.3$ are due to the 4000 \textup{\AA} break. We again use the correction derived from stars with the same $g-i$ and the SCE residuals after correction are shown as red dashed lines in the same figure.  

Similar to SNe Ia, the SCE are not properly corrected since the galaxy SED is different from that of a star. However, it is worth noting that the star-derived corrections using $g-i$ color actually slightly correct the SCE on galaxies. For example,  in $griz$-bands, the red lines (after correction) are much closer to zero compared to the blue lines (before correction) for redshift $z<1$. Here, we derive the corrections based on a fixed color $g-i$. This might be good enough for regular stars, as they form a well-defined stellar locus on a color-color diagram. We pick $g-i$ throughout the paper since $g-i$ ranging from $-1$ to $4$ separates the blue stars from red stars\footnote{For example, $g-r$ is not a good color to pick for red stars since all K and M stars tend to clump around $g-r\sim1.5$. Therefore, it is hard to correct the SCE on red stars using $g-r$ color.}. However, $g-i$ might not be the best choice for SNe Ia and galaxies. Choosing a color close to the band of interest, e.g., $g-r$ for $g$-band, and $z-Y$ for $z$-band and/or $Y$-band, or choosing a combination of multiple colors, might be better for these non-stellar objects. We leave further discussion of this to future work.


\begin{table*}[th!]
\small
\centering
\caption{Peak-to-Valley SCE on SNe Ia over redshift $z=0$ to $z=1$.}
\begin{tabular}{c|c|c|cccccc}
\hline
\multirow{2}{*}{Component} & \multirow{2}{*}{Fiducial} & \multirow{2}{*}{Changed} & \multicolumn{6}{c}{Synthetic SCE (mmag)} \\ \cline{4-9} 
                      &                     &                     & u  & g  & r  & i  & z  & Y \\ \hline
Pressure & $P_0=779$ hpa    & $P_0=789$ hpa & 0.8 & 0.9 & 0.1 & 0.2 & 0.1 & 0.1 \\ \hline
\multirow{2}{*}{Aerosol}     & AOD$_{550}=0.02$, $\alpha=1$ & AOD$_{550}=0.20$, $\alpha=1$  & 11 & 26 & 7 & 4 & 2 & 1 \\ \cline{2-9} 
           & AOD$_{550}=0.02$, $\alpha=1$ & AOD$_{550}=0.20$, $\alpha=0.5$  & 4 & 11 & 3 & 2 & 1 & 0.7 \\ \hline
 PWV       & PWV $=3$ mm    & PWV $=10$ mm &  0 &  0 & 3 & 7 & 20 & 27\\ \hline

 Ozone     & Ozone $=270$ DU & Ozone $=230$ DU & 0.04 & 1.7 & 1 & 0.1 & 0 & 0 \\ \hline
 Airmass   & $X=1.2$      & $X=1.8$ & 31 & 30 & 7 & 4 & 6 & 7 \\ \hline
 Instrument & DECal scan   & shift 2 nm & 186 & 99 & 33 & 22 & 19 & 28 \\ \hline
\end{tabular}

\label{table:sn_error}
\end{table*}

\begin{table*}[th!]
\small
\centering
\caption{Peak-to-Valley SCE on elliptical galaxies over redshift $z=0$ to $z=2$.}
\begin{tabular}{c|c|c|cccccc}
\hline
\multirow{2}{*}{Component} & \multirow{2}{*}{Fiducial} & \multirow{2}{*}{Changed} & \multicolumn{6}{c}{Synthetic SCE (mmag)} \\ \cline{4-9} 
                      &                     &                     & u  & g  & r  & i  & z  & Y \\ \hline
Pressure & $P_0=779$ hpa    & $P_0=789$ hpa & 0.6 & 0.6 & 0.07 & 0.1 & 0.04 & 0.03 \\ \hline
\multirow{2}{*}{Aerosol}     & AOD$_{550}=0.02$, $\alpha=1$ & AOD$_{550}=0.20$, $\alpha=1$  & 7 & 18 & 10 & 4 & 3 & 1 \\ \cline{2-9} 
           & AOD$_{550}=0.02$, $\alpha=1$ & AOD$_{550}=0.20$, $\alpha=0.5$  & 3 & 7 & 5 & 2 & 2 & 0.7 \\ \hline
 PWV       & PWV $=3$ mm    & PWV $=10$ mm &  0 &  0 & 4 & 4 & 26 & 20\\ \hline

 Ozone     & Ozone $=270$ DU & Ozone $=230$ DU & 0.03 & 1.2 & 1.6 & 0.1 & 0 & 0 \\ \hline
 Airmass   & $X=1.2$      & $X=1.8$ & 26 & 21 & 11 & 4 & 7 & 6 \\ \hline
 Instrument & DECal scan    & shift 2 nm & 79 & 36 & 28 & 26 & 17 & 22 \\ \hline
\end{tabular}

\label{table:elgal_error}
\end{table*}

\section{Systematic Chromatic Errors In Ground-based Transit Observations}
We discussed the SCE in this paper mainly for the large area sky surveys like DES. However, the SCE calculated here are not only limited to photometric calibrations in large surveys. The synthetic calculations presented in Table \ref{table:star_error} can also be applied to any ground-based differential photometric measurements, such as exoplanet transients and other variable star measurements. For example, if the planet host is an M6V star and most of the reference stars in the field are G2V stars, then PWV varying from 3~mm to 10~mm can affect the photometry in $z$-band to $\sim$ 1\%. This is comparable to or even larger than the signal from a super-Earth transiting an M star. Similarly, variations in either the atmospheric transmission or instrumental throughput can also affect such measurements in other bands. If we correct the SCE in these differential measurements with auxiliary systems, it is possible to improve the detection of Earth-like exoplanets around M star hosts in some ground-based transit observations such as the MEarth Project~\citep{Irwin2009, Berta-Thompson2015}.

\section{Conclusions}
In this paper, we have demonstrated that the variation of the atmospheric transmission and the instrumental throughput introduce systematic chromatic errors (SCE) that depend on the color of the source object. We assess such SCE for the Dark Energy Survey (DES) as an example: 
\begin{itemize}
\item For stars, the SCE caused by the change of airmass in each exposure and the change of precipitable water vapor and aerosol in the atmosphere can be larger than 1\%;
\item The SCE caused by the change of the barometric pressure and ozone are smaller than 0.1\% (or 1 mmag); 
\item The SCE caused by the bandpass edge shift over the detector focal plane can be as large as a few percent.
\item The SCE can be corrected to 2-3 mmag or better if the shape of the atmospheric transmission and the instrumental throughput are well measured by auxiliary calibration system such as aTmCam and DECal.  
\item For supernovae and galaxies, these SCE are expected to be larger and also redshift-dependent. Figure \ref{fig:error_pwv_sn} and \ref{fig:error_bandpass_shift_elgal} give examples of how SCE change as a function of redshift for Type Ia supernovae and elliptical galaxies.
\item For stars, we could derive a color term to first order and approximately correct the SCE. However, such linear stellar color terms are not sufficient for getting color corrections for extremely red stars, SNe and galaxies. 

\end{itemize}
From this study, we suggest that, for large imaging surveys such as DES and LSST, one should first define a natural system response that represents the average condition of the survey, and also define a stellar SED as the reference SED that represents the average color of the calibration stars in the survey. Then for each stellar object, one could calculate the synthetic SCE using the stellar SED library together with the stellar color, as described in this paper, and use the synthetic SCE as the corrections. For non-stellar objects like SNe Ia or galaxies, the redshifted SED need to be given as an input to derive such corrections.

Even though SCE are systematic errors for each exposure, they will eventually introduce additional scatter on the final coadd photometry and affect the photometric precision of the surveys with multiple visits. This is true because: 1) the atmospheric transmission varies over a wide range of conditions and each exposure is likely to be taken with a different condition; 2) each exposure in the survey has a slight offset for multiple tilings so that the same object does not fall on the same location on the focal plane. Averaging over different conditions and different focal plane positions can reduce the amount of SCE in the final coadd photometry.  Exceptions include supernovae and other transients, for which there is only one measurement in a particular epoch.

\acknowledgements
This paper has gone through internal review by the DES collaboration. We thank
the anonymous referee for comments and suggestions that improved the paper. TSL thanks Michael Smitka, Nicholas Sunzeff, {\v Z}eljko Ivezi{\'c} and Abhijit Saha for very helpful conversations.
Texas A \& M University thanks Charles R. '62 and Judith G. Munnerlyn, George P. '40 and Cynthia Woods Mitchell, and their families for support of astronomical instrumentation activities in the Department of Physics and Astronomy.

Funding for the DES Projects has been provided by the U.S. Department of Energy, the U.S. National Science Foundation, the Ministry of Science and Education of Spain, 
the Science and Technology Facilities Council of the United Kingdom, the Higher Education Funding Council for England, the National Center for Supercomputing 
Applications at the University of Illinois at Urbana-Champaign, the Kavli Institute of Cosmological Physics at the University of Chicago, 
the Center for Cosmology and Astro-Particle Physics at the Ohio State University,
the Mitchell Institute for Fundamental Physics and Astronomy at Texas A\&M University, Financiadora de Estudos e Projetos, 
Funda{\c c}{\~a}o Carlos Chagas Filho de Amparo {\`a} Pesquisa do Estado do Rio de Janeiro, Conselho Nacional de Desenvolvimento Cient{\'i}fico e Tecnol{\'o}gico and 
the Minist{\'e}rio da Ci{\^e}ncia, Tecnologia e Inova{\c c}{\~a}o, the Deutsche Forschungsgemeinschaft and the Collaborating Institutions in the Dark Energy Survey. 
The DES data management system is supported by the National Science Foundation under Grant Number AST-1138766.
The DES participants from Spanish institutions are partially supported by MINECO under grants AYA2012-39559, ESP2013-48274, FPA2013-47986, and Centro de Excelencia Severo Ochoa SEV-2012-0234, 
some of which include ERDF funds from the European Union.

The Collaborating Institutions are Argonne National Laboratory, the University of California at Santa Cruz, the University of Cambridge, Centro de Investigaciones En{\'e}rgeticas, 
Medioambientales y Tecnol{\'o}gicas-Madrid, the University of Chicago, University College London, the DES-Brazil Consortium, the University of Edinburgh, 
the Eidgen{\"o}ssische Technische Hochschule (ETH) Z{\"u}rich, 
Fermi National Accelerator Laboratory, the University of Illinois at Urbana-Champaign, the Institut de Ci{\`e}ncies de l'Espai (IEEC/CSIC), 
the Institut de F{\'i}sica d'Altes Energies, Lawrence Berkeley National Laboratory, the Ludwig-Maximilians Universit{\"a}t M{\"u}nchen and the associated Excellence Cluster Universe, 
the University of Michigan, the National Optical Astronomy Observatory, the University of Nottingham, The Ohio State University, the University of Pennsylvania, the University of Portsmouth, 
SLAC National Accelerator Laboratory, Stanford University, the University of Sussex, and Texas A\&M University.


\bibliographystyle{apj}
\bibliography{bib}

\end{document}